\documentclass[11pt]{article}
\usepackage{comment}
\usepackage[hidelinks]{hyperref}
\usepackage{enumitem}
\usepackage{amsmath,amssymb,epsf,cite,graphicx,subfigure}
\usepackage{amsmath,amsthm, tikzsymbols, dsfont, amsfonts}
\usepackage{comment}
\usepackage{xcolor}
\usepackage{graphicx}
\usepackage{dsfont}
\usepackage{soul}
\usepackage{mathrsfs}
\usepackage[export]{adjustbox}

\numberwithin{equation}{section}

\definecolor{airforceblue}{rgb}{0.36, 0.54, 0.66}
\newcommand{\beq}{\begin{equation}}
\newcommand{\eeq}{\end{equation}}

  \theoremstyle{definition}

 \def\ee{\end{equation}}
\def\be{\begin{equation}}
\def\bea{\begin{eqnarray}}
\def\eea{\end{eqnarray}}
\newcommand{\eqq}{\end{eqnarray}}
 \newcommand{\badat}{\begin{alignedat}}
 \newcommand{\eadat}{\end{alignedat}}

\newcommand{\eal}[1]{\be \begin{aligned} #1 \end{aligned}\end{equation}} 
\newcommand{\eqn}[1]{\be #1 \end{equation}} 
\newcommand{\eqa}[1]{\bea  #1\end{eqnarray}}

\long\def\new#1\endnew{{\bf #1}}		
\long\def\del#1\enddel{}

\def\del{\partial}

 \def\p{\partial }
 \def\cm{{\cal I}^-_R}
 \def\cp{{\cal I}^+_R}

\def\cp{{\cal I}^+_R}
\def\cm{{\cal I}^-_R}

\begin{document}
\baselineskip=15.5pt
\pagestyle{plain}
\setcounter{page}{1}
%--------+---------+---------+---------+---------+---------+---------+

\begin{center}
{\LARGE \bf %Unitarity in Time-dependent Cosmologies
% Unitarity in Expanding Universes
% The Universe as a Quantum Encoder
% The Expanding Universe as a Quantum Embedding
% The Expanding Universe as a Quantum Encoder
The Universe as a Quantum Encoder 
%The Expanding Universe as a Quantum Encoder
}
\vspace{0.5cm}

\textbf{Jordan Cotler and Andrew Strominger}

\vspace{0.4cm}
{\it Society of Fellows, Black Hole Initiative and Department of Physics \\}
{\it Harvard University, Cambridge, MA 02138 USA \\}

%{\it Harvard Society of Fellows, Cambridge, MA 02138 USA \\}
%{\it Black Hole Initiative, Harvard University, Cambridge, MA 02138 USA \\}
%{\it Center for Fundamental Laws of Nature, Harvard University, Cambridge, MA 02138 USA \\}

\vspace{0.1cm}

\vspace{.1cm}

\end{center}

\begin{center}
{\bf Abstract}
\end{center}
Quantum mechanical unitarity in our universe is challenged both by the notion of the big bang, in which nothing transforms into something, and the expansion of space, in which something transforms into more something. This motivates the hypothesis that quantum mechanical time evolution is always isometric, in the sense of preserving inner products, but not necessarily unitary.  As evidence for this hypothesis we show that in two spacetime dimensions (i) there is net entanglement entropy produced in free field theory by a moving mirror or expanding geometry, (ii) the Lorentzian path integral for a finite elements lattice discretization gives non-unitary isometric time evolution, and (iii) tensor network descriptions of AdS$_3$ induce a non-unitary but isometric time evolution on an embedded two-dimensional de Sitter braneworld.  In the last example time evolution is a quantum error-correcting code. 

\newpage

\tableofcontents

\newpage

\section{Introduction}

The following three long-cherished beliefs about the physical universe,
\begin{enumerate} \item Quantum states on different  time slices are related  by unitary transformations; 
\item The universe is expanding;
\item There are no degrees of freedom on length scales shorter than the Planck scale;
\end{enumerate} 
are in considerable tension with one another. (2) and (3) together strongly suggest the dimension of the Hilbert space grows with time. This precludes the unitary maps of (1), since they can exist only if the Hilbert spaces are the same size.\footnote{This tension is also at the core of the black hole information paradox in that Hawking quanta descend from super-Planckian modes in an expanding local geometry \cite{PhysRevD.44.1731}.}  Alternately the two beliefs
\begin{enumerate} \item Quantum states on different  time slices are related  by unitary transformations;
\item The universe originated in a big bang before which there was nothing; \end{enumerate} 
are also in tension with one another because there is no unitary transformation from nothing to something. The nature of this second tension, although obviously related to the first one,  is different because it entails consideration of a region of spacetime controlled by unknown  strongly coupled quantum gravitational dynamics.  Therefore we focus on the first tension between (1)-(3) about which more can be said. 

In this paper we explore  the radical idea that in our universe the tension is resolved by relaxing the condition (1) of unitarity. We hypothesize that time evolution is always isometric but not necessarily unitary. Isometries are maps from a smaller to a larger Hilbert space which preserve the inner product between any two states. If the Hilbert spaces have the same size, they become unitary transformations. In an expanding universe they allow new degrees of freedom to be added without affecting inner products, giving mathematical precision to the intuitive statement that ``new degrees of freedom appear in their ground state" \cite{Polchinski:1995ta, Strominger:1994tn}\footnote{Polchinski comes quite close to the picture that we present here, saying that low-energy time evolution is a ``one-way unitary". But he does not abandon microscopic unitary, or specify how the newly sub-cutoff modes are to be entangled with other modes.} and specifying their quantum correlations with the preexisting degrees of freedom.  Isometries play a central role in instantiating encodings in quantum information theory: here we take a further step in suggesting a fundamental role in physical law.  Among all isometries, of special interest are ones instantiating quantum error-correcting codes which are related to renormalization group (RG) flow and holography \cite{Almheiri:2014lwa,Pastawski:2015qua, Jahn:2021uqr}. 

We motivate this hypothesis from three different approaches, all in 1+1 dimensions:  a continuum analysis of entanglement entropy, latticization of quantum field theory on an expanding geometry using the finite elements method, and a braneworld embedding of dS$_2$ in the holographic formulation of AdS$_3$ modeled by a tensor network quantum error-correcting code. 

The first, very simple, example is a 1+1 free field theory with a moving mirror~\cite{fulling1976radiation, Wilczek:1993jn, Fiola:1994ir} (see~\cite{Akal:2021foz} for more recent work).  We take the mirror to initially be at rest, and then uniformly accelerating away from infinity.  The mirror then stops accelerating and remains at a final receding velocity. Since the volume of spatial slices increases, this serves as a toy model of an expanding universe.\footnote{And also of black hole evaporation.} This problem is analytically soluble. The entanglement entropy of the outgoing radiation rises thermally during the acceleration and then remains at a nonzero constant: there is no Page-type return to zero.  An observer at infinity with any finite UV cutoff will never see the thermal radiation purified no matter how long they wait. This is due to the fact that some super-cutoff incoming modes are redshifted below the cutoff by the receding mirror before arriving at the observer. Hence some modes which could not contribute to entanglement entropy prior to the acceleration may do so afterward.  Because new effective degrees of freedom have emerged, this is described by an isometric but non-unitary transformation. This simple continuum example illustrates the fact that non-unitary isometries arise no matter how high we place the cutoff. We expect that any discretization which has a good continuum limit will give isometries and not unitaries simply because the dimension of the Hilbert space must grow. 

The second approach is the lattice in which a finite UV cutoff is manifest.  Lattice discretizations of Euclidean quantum field theory (QFT) in curved space
have been studied in\cite{Brower:2016vsl, Brower:2018szu, Brower:2019kyh, Brower:2016moq} and are generally best achieved using (variants) of the finite elements method (FEM)~\cite{ern2004theory, brenner2008mathematical}, which we review.  Our contribution is to formulate and study the Lorentzian setting, which has some remarkable properties.  Many lattice realizations of QFT in an expanding geometry necessitate adding lattice points, and a unitary transformation between time slices is clearly impossible. We construct the Lorentzian FEM path integral\footnote{The more familiar finite difference method (FDM) used in most QFT applications is not flexible enough for general curved spacetimes. It breaks down already at the classical level for the moving mirror and does not have a good continuum limit. } and show explicitly that time evolution between subsequent slices is an isometry. This demonstrates that discrete path integral methods are not restricted to unitary systems and naturally describe the more general isometries of interest to us here.  For related work in the canonical quantization setting, see~\cite{Foster:2004yc, Dittrich:2013jaa, Hohn:2014uvt, Hohn:2014rba}.

Our third approach leverages beautiful work on AdS$_3$/CFT$_2$  holography \cite{Almheiri:2014lwa,Pastawski:2015qua, Jahn:2021uqr},
 where {\it radial} evolution at fixed time in AdS$_3$ was modeled by a tensor network quantum error-correcting code. 
 This is intimately tied to the fact that radial evolution is inverse RG flow, which is known in general to be an approximate error-correcting code~\cite{Kim:2016wby, Furuya:2020tzv, Furuya:2021lgx}.
This work is combined with the construction of quantum gravity in de Sitter space as a braneworld near the boundary of AdS~\cite{Hawking:2000da}. de Sitter expansion involves radial motion in AdS. Putting these together within the tensor network model leads to de Sitter time evolution as a quantum error-correcting code. 
Instantiating this within full-fledged AdS/CFT would be a refinement of and imply the dS/CFT correspondence in which time evolution is holographically dual to RG flow~\cite{Strominger:2001pn,Strominger:2001gp}.  Our work more broadly suggests that toy tensor network models of de Sitter with isometric time evolution~\cite{SinaiKunkolienkar:2016lgg, Bao:2017qmt, Milsted:2018san, Niermann:2021wco} should be taken more seriously as capturing properties of time evolution in the real world.

% which instantiate isometries as time evolution may share in common features with our own universe. 
% a refinement of the dS/CFT correspondence in which time evolution is holographically dual to RG flow~\cite{Strominger:2001pn}. 
% These observations are a refinement of and imply the dS/CFT correspondence in which time evolution is holographically dual to RG flow~\cite{Strominger:2001pn}. 

This paper is organized as follows.  In section~\ref{Sec:EE} we study the entanglement of a 2D CFT in two kinds of expanding cosmologies: flat space with a receding mirror boundary, and a closed expanding universe.  In section~\ref{Sec:Lattice} we review the finite elements method and explain its necessity for discretizing classical equations of motion for field in curved backgrounds. We then perform a Lorentzian path integral quantization and show that time evolution is isometric.  In section~\ref{Sec:embedding} we interface the dS$_2$ braneworld proposal with a tensor network quantum error correction model of AdS$_3$/CFT$_2$.  In section~\ref{Sec:dSCFT} we explain implications of our work to dS/CFT.  Section~\ref{Sec:Discuss} contains some closing comments including a remark about contracting geometries.  The appendices contain additional technical details about our Lorentzian FEM path integrals, and how they instantiate isometries as time evolution.

\section{2D CFT entanglement entropy }
\label{Sec:EE}
\subsection{Receding  mirrors}
\begin{figure}
\begin{center}
\includegraphics[scale=.5]{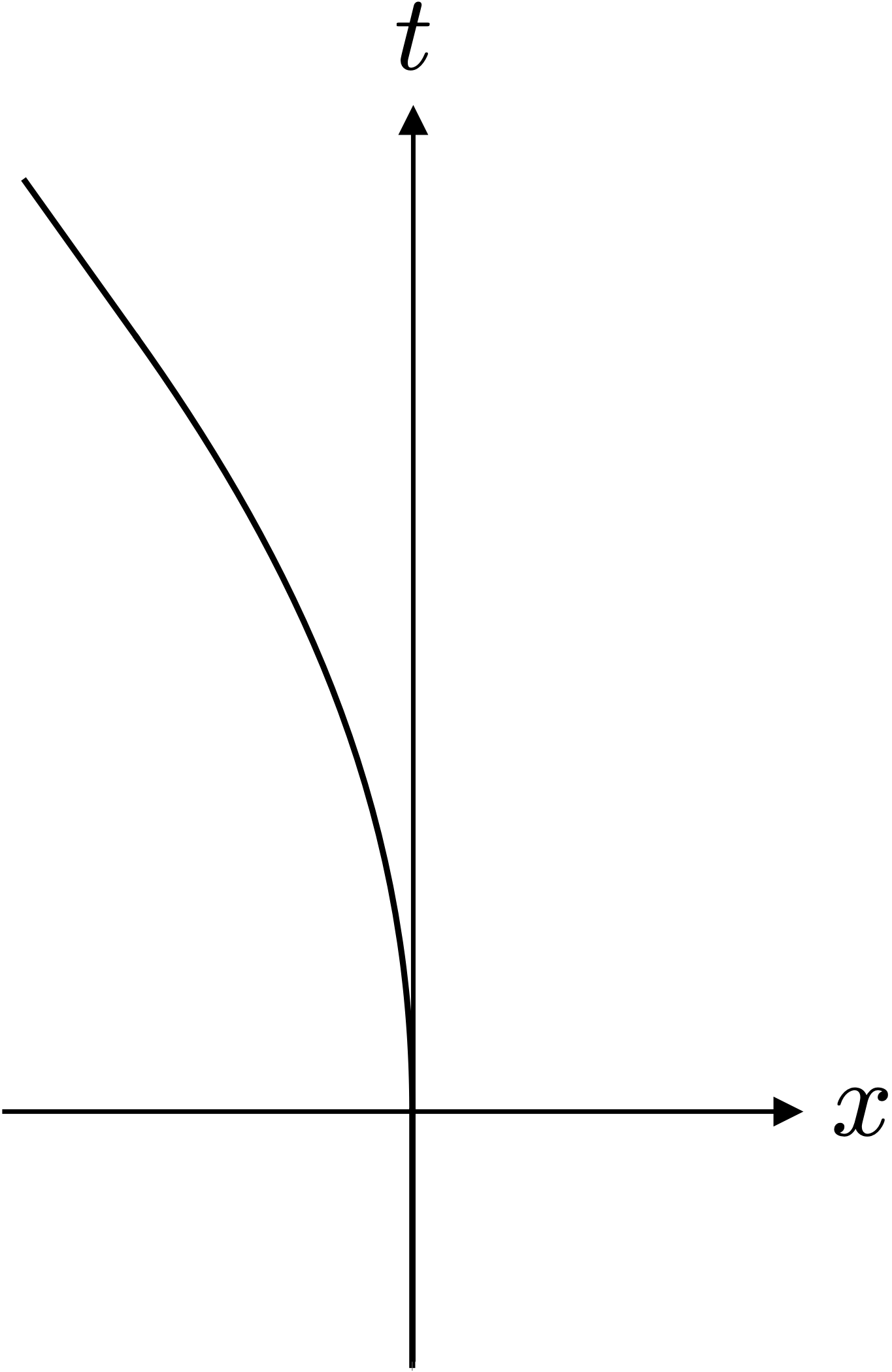}
\end{center}
\caption{A mirror following the trajectory in Equation~\eqref{pol}. \label{Fig:MovingMirror1}}
\end{figure}

Two dimensional Minkowski space has the metric
\be\label{ffl} ds^2=-dt^+dt^-\,.\ee
We refer to these as `inertial coordinates' since inertial trajectories are of the form $t^+=a\,t^-+b$. 
Here $t^-$ ($t^+$) is a coordinate on $\cp$ ($\cm$). 
We wish to consider a free scalar in the presence of a mirror so there is no ${\cal I}^\pm_L$. To model black hole formation/evaporation we take the incoming state to be the vacuum on $\cm$ and the  mirror to be initially stationary; then the mirror accelerates away from infinity for a retarded time $L$ on $\cp$, and then moves inertially as depicted in Figure~\ref{Fig:MovingMirror1}. This also models an expanding universe because the volume of space is increasing. 
The mirror trajectory is described by $t^+=\tilde t^-(t^-)$ with 
\bea \label{pol}\tilde t^-&=&t^-,~~~~t^-<0\,,\cr &=&{1 \over 2 \pi T_H}\bigl(1-e^{-2\pi T_Ht^-}\bigr)\,,~~~~0<t^-<L\,,\cr &=&e^{-2\pi T_H L}(t^--L)+{1 \over 2\pi T_H}\bigl(1-e^{-2\pi T_HL}\bigr)\,,~~~~t^->L\,.\eea Note that the final inertial trajectory is boosted relative to the initial one. The 
 outgoing state, which is a wavefunction  on $\cp$, is given by the vacuum state with respect to the `vacuum coordinates' 
$(t^+,\tilde t^-)$.   In vacuum  coordinates the metric \eqref{ffl} is 
\be \label{dsc} ds^2= -e^{2\rho}d\tilde t^-dt^+,~~~~~\rho=-{1 \over 2} \log \p_- \tilde t^- .\ee
Inertial detectors do not move along lines of constant $\tilde t^--t^+$ and so will detect particles. In the intermediate accelerating phase,  $\tilde t^-$ is invariant under $t^- \to t^- +{i \over T_H}$ so the outgoing flux appears thermal at temperature $T_H$.  Outside the time period $0<t^-<L$ the mirror is inertial and there is no flux at infinity. This resembles a black hole radiating for a retarded time $L$. However the outgoing state is obviously pure.

An observer measuring the portion of the quantum state on $\cp$  prior to some fixed finite retarded time $t^-$ will generically find a mixed state. This is characterized by the  entanglement entropy  $S_{ent}(t^-)$ of the portions of the quantum state on $\cp$ before and after $t^-$. 
This entanglement has a UV divergent\footnote{A recent discussion of the IR divergences can be found in \cite{Akal:2021foz}.} constant ($t^-$ independent) piece which we subtract so that it vanishes at early times $t^-\to-\infty$. Real detectors cannot measure arbitrarily short wavelengths so this subtracted finite quantity is the actual observed entropy of the portion of the (effectively mixed) quantum state measured prior to $t^-$.  For a conformal field theory of central charge $c$, the result is \cite{Holzhey:1994we}
\be \label{ddd}S_{ent}(t^-)={c \over 6}\, \rho(t^-) \,,\ee
where $\rho(t^-)$ is given in~\eqref{dsc}. (For a free scalar $c=1$.) 
One finds for \eqref{dsc}
\bea \label{snt} S_{ent}&=&0\,,~~~~~~~~~~~~t^-<0\,,\cr  &=& {\pi cT_Ht^- \over 6}\,,~~~~0<t^-<L\,,\cr &=&{\pi cT_HL \over 6}\,,~~~~t^->L\,.\eea
This can be compared to the formula for the energy flux \be \label{efl}T_{--}={c \over 24\pi}\,e^{-\rho}\p_-^2e^{\rho} \ee at $\cp$ where
\bea T_{--}&=&0\,,~~~~~~~~~~~~t^-<0\,,\cr  &=& {c\pi T_H^2 \over 12}\,,~~~~0<t^-<L\,,\cr &=&0\,,~~~~~~~~~~~~t^->L\,,\eea
which is thermal during the acceleration. 
Hence the entanglement entropy is zero prior to acceleration, increases as expected for thermal radiation at temperature $T_H$ during acceleration, and then remains constant afterward. 

Perhaps surprisingly, the limit as the entanglement point $t^-$ goes to future timelike infinity is not trivial.  Nevertheless the state on $\cp$ is manifestly pure. This is possible because of the ambiguous constant associated to the UV divergences in the entanglement entropy. We have to throw away the entanglement across $t^-$ of modes with proper wavelength shorter than the UV cutoff. However after the acceleration, the mirror is receding and modes above the cutoff from $\cm$ are Doppler shifted into the IR below the cutoff before they reach $\cp$. Therefore some wavelengths on $\cm$ which do not contribute to the entanglement on $\cp$ in the early period do contribute in the late period, and the subtraction constant is shifted. 

Interestingly, an experimentalist stationed at $\cp$, uninformed of these subtleties, will never see the purity of the quantum state restored for any finite $t^-$ and will conclude that information is destroyed. This is ultimately due to the fact that entanglement entropy is not a well-defined `number' and is beset by divergences which must be subtracted. (This is potentially relevant to the black hole information paradox \cite{Kapec:2016aqd}.)

In this example the effective Hilbert space -- composed of modes below the cutoff -- is larger on $\cp$ than on $\cm$ as  a consequence of  the expansion of the bulk space and associated  new degrees of freedom.  Hence there is no effective unitary description no matter how high  the UV cutoff. We will argue  in the following sections that the map from the in to the out Hilbert space can be effectively described by an isometry. 

\subsection{Cosmology}
Consider a free scalar in a closed $1+1$ cosmology  in conformal coordinates
\be\label{cc} ds^2=-R^2(t)\,dt^+dt^-\,,~~~t^\pm=t\pm \sigma, \,\,\,\sigma \sim\sigma+2\pi\,.\ee
If the radius $R$ goes to a constant in the far past, $t^\pm$ are vacuum coordinates defining a vacuum in which inertial detectors in the far past see no particles.  We are interested in the entanglement entropy between the states on the left and right side of the circle, dividing the circle in half at  $t^+-t^-=0,2\pi$ and setting the constant to give zero in the far past.  Equation \eqref{ddd} in fact holds also in curved space provided $\rho$ is taken to be the conformal factor in vacuum coordinates \cite{Fiola:1994ir}. This gives simply 
\be \label{rgg} S_{ent}={c\over 3} \log {R(t)\over R(-\infty)}\,. \ee
Hence, as the universe expands, the entanglement entropy grows. We interpret this as originating from new entangled degrees of freedom which are redshifted below the cutoff by the expansion. 

Now let us consider de Sitter space.  Planar coordinates
\be ds^2=-{\ell^2\over (u^++u^-)^2}\,du^+du^-, ~~~~~u^\pm=u\pm x\,,\ee
with no identification on $x$, cover a future expanding  half of de Sitter space with flat $u=$ constant slices.
These are vacuum coordinates which define the Hartle-Hawking  vacuum.  Consider a subregion with fixed unit coordinate length in $x$ and on a line of constant $u$. The entanglement entropy of the vacuum state restricted to 
that region is 
\be\label{dse} S_{ent}={c \over 3}\,\log {\ell\over u}\,.\ee
 Defining $u=e^{-t}$ it grows linearly with time $t$.  As with the moving mirror, we can understand this as arising from the growth of the effective Hilbert space from the redshifting cutoff.

\section{Unitarity and isometry on the lattice}
\label{Sec:Lattice}

Lattice quantum field theory provides a UV regularization of continuum field theory which has had extraordinary success in numerical applications, most notably lattice QCD (see~\cite{creutz1983quarks} for an accessible overview of the formalism).  While much attention has been given to providing suitable lattice discretizations of fermions and gauge fields, comparatively less attention has been given to Lorentzian field theory on time-dependent backgrounds.  We pursue this here, initially in the context of free scalar field theory.

On the one hand, constructing a lattice discretization of a scalar field theory seems disarmingly intuitive; we might be inclined to examine the action $S$ and simply promote derivatives to finite differences entailing a lattice scale.  But in time-dependent backgrounds, the task has interesting subtleties.  Before delving into subtleties at the quantum level, we should begin with an essential, classical consideration: we must choose a lattice discretization of the action such that in the appropriate continuum limit the action principle recovers the continuum classical equations of motion.  Accordingly, we should understand how to appropriately discretize a PDE on a lattice.

\subsection{Classical considerations}

The most straightforward manner of discretizing a PDE is via the finite difference method (FDM).  We will review this first; it is particularly simple, and as such was historically the first discretization method to be discovered.  However, the FDM has fatal deficiencies for PDE's on certain kinds of time-dependent backgrounds, as we will discuss.  These issues can be resolved using the finite elements method (FEM) which leverages comparatively more sophisticated tools from functional analysis~\cite{ern2004theory, brenner2008mathematical}.   FEM and its variants are the workhorses of modern numerical PDE solvers~\cite{liu2021eighty}. We provide a review of this method in subsection 3.1.2 aimed towards our application to lattice field theory in expanding geometries. 

\subsubsection{Finite difference method}

\begin{figure}
\begin{center}
\includegraphics[scale=.5]{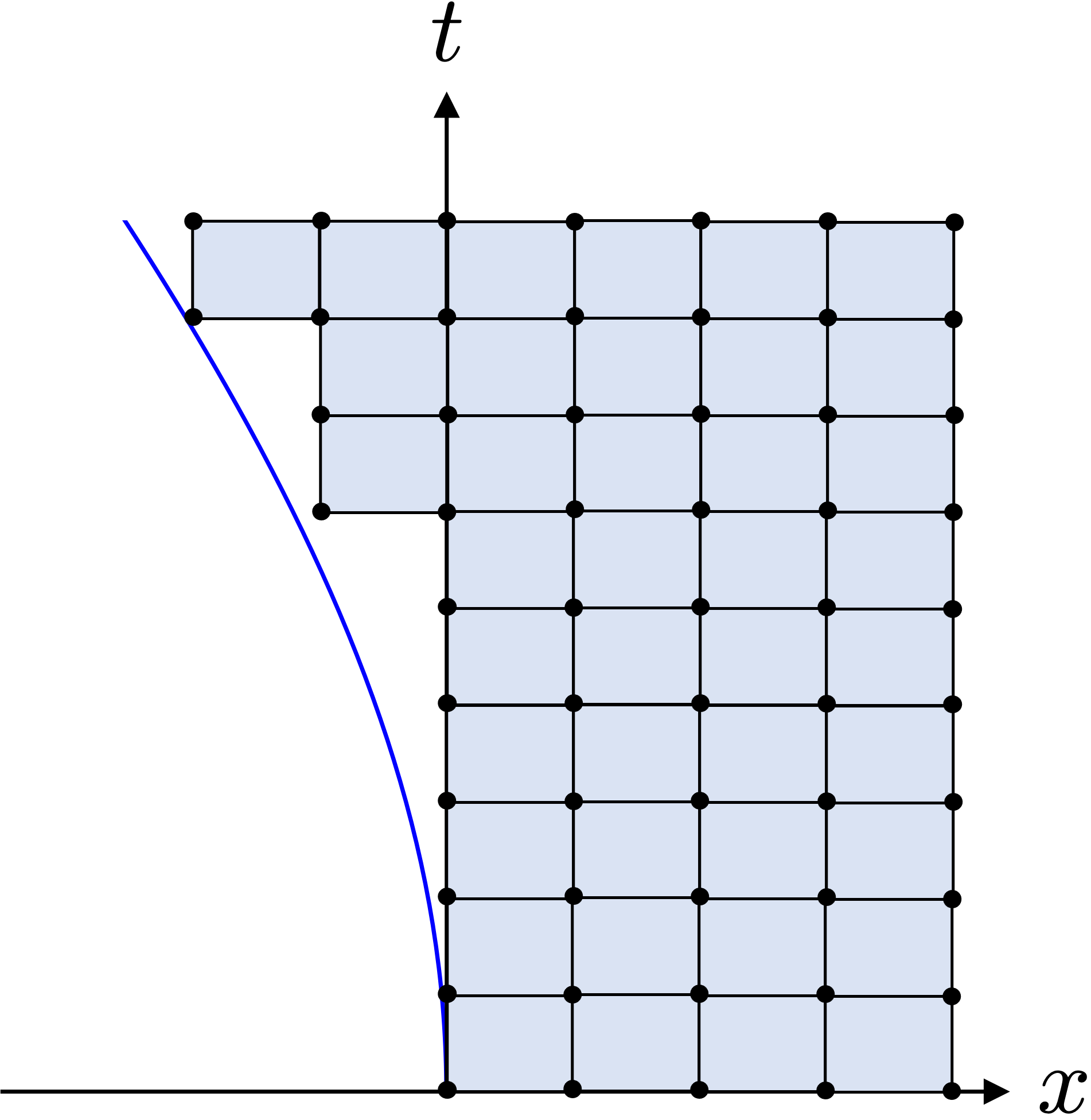}
\end{center}
\caption{An FDM rectangular lattice regularization for a 1+1 spacetime with a moving left boundary.  This was obtaining by tiling the entire upper half-plane with rectangles, and removing those rectangles which intersect with or fall to the left of the moving mirror boundary.  To impose Dirichlet boundary conditions on the mirror, the field degrees of freedom on the left-most lattice sites of the remaining lattice are set to zero.  A wave reflected off of the saw-tooth boundary
%becomes nondifferentiable and has 
can acquire an energy flux which becomes large in the $a\to 0$ continuum limit.     \label{fig:timedepbdy2}}
\end{figure}

Let us consider a 1+1 free massless scalar field in flat space:
\begin{equation}
S[\phi] = -\frac{1}{2}\int d^2 x \, \partial_\mu \phi \,\partial^\mu \phi \,.
\end{equation}
The FDM approach to discretization is to regularize according to a rectangular spacetime lattice, where spatial lattice links have length $\delta x$ and temporal lattice links have length $\delta t$.  We take $\delta x, \delta t \propto a$ which we call the lattice spacing.  From an EFT point of view, we should regard the minimum value of $a$ as being an $O(1)$ multiple of the Planck length $\ell_P$, say $a = 100 \,\ell_P$.  When we speak of $a \to 0$, we really mean that $a$ is going to $\sim 100 \,\ell_P$.  More generally, we can have a rectangular lattice where the lattice spacing varies depending on the location in spacetime (e.g.~Figure~\ref{fig:timedepbdy2}).   Then we promote $\phi(t,x)$ to $\phi_{i,j}$ which is labeled by the vertices $i,j$, and further promote the derivatives to finite differences and the integral to a finite sum so that we obtain
\begin{equation}
\label{E:latticeaction}
S_{\text{lattice}}[\{\phi_{i,j}\}] = -\frac{1}{2} \sum_{i,j} \delta t \, \delta x \,\eta^{\mu \nu}\Delta_\mu \phi_{i,j} \,\Delta_\nu \phi_{i,j} \,.
\end{equation}
Here we have
\begin{equation}
\Delta_0 \phi_{i,j} = \frac{\phi_{i+1,j} - \phi_{i,j}}{\delta t}\,, \qquad \Delta_1 \phi_{i,j} = \frac{\phi_{i,j+1} - \phi_{i,j}}{\delta x}\,.
\end{equation}
%Above, we have assumed that the metric $g_{\mu \nu}(t,x)$ is sufficiently regular at the lattice scale, i.e. that the lattice scale is much smaller than the typical distance scale of changes in the metric.  Suppose for simplicity that $g_{\mu \nu}$ is diagonal.
Then varying $S_{\text{lattice}}[\{\phi_{i,j}\}]$ with respect to $\phi_{i,j}$ we find the lattice equations of motion,
\begin{equation}
- \frac{2 \phi_{i,j} - \phi_{i+1,j} - \phi_{i-1,j}}{\delta t^2} + \frac{2 \phi_{i,j} - \phi_{i,j+1} - \phi_{i,j-1}}{\delta x^2} = 0\,,
\end{equation}
where the first and second terms implement a discrete Laplacian in the time and space directions, respectively.  Since $\delta t, \delta x \propto a$, as $a \to 0$, this becomes the usual Klein-Gordon equation.\footnote{For purposes of numerical stability of the classical equations, it is useful to let $\frac{\delta t}{\delta x} < 1$.  This ratio $\frac{\delta t}{\delta x}$ is often called the Courant number~\cite{courant1967partial}.  For a massive theory, this condition can be relaxed.}

The FDM is the standard approach of lattice quantum field theory~\cite{creutz1983quarks}, usually implemented in Euclidean signature.  A lattice action of the form~\eqref{E:latticeaction}, often with added interactions, is quantized via the path integral.  However, there are a large number of circumstances for which the FDM is unsuitable, even for reproducing the correct \textit{classical} equations of motion in the $a \to 0$ limit.  These are not circumstances ordinarily considered in lattice field theory, but are highly relevant to our present considerations.  For example, FDM typically fails if we try to propagate a field in a curved or time-dependent background or even with a curved or time-dependent boundary. Perhaps the simplest example is a free 1+1 scalar field on an interval with Dirichlet boundary conditions where the location of the left-most boundary is time-dependent, as considered in section 2.1.  See Figure~\ref{fig:timedepbdy2} for a depiction.  The saw-tooth approximation of the left boundary is inadequate for an appropriate continuum limit: even as we take the lattice spacing to zero, the limiting curve of the left boundary will not be differentiable. A classical wave reflected off the saw-tooth with Dirichlet boundary conditions can acquire an energy flux that blows up in the $a \to 0$ continuum limit.
%At the quantum level, this effect At the quantum level even the incoming vacuum acquires divergent energy: the energy flux \eqref{efl} has an unsupressed a delta-function squared contribution 
% emanating from the corner of each sawtooth.  Hence the sawtooth boundary injects Planckian amounts of energy into the interior. 

Similar problems occur when coupling fields to a time-dependent cosmology, such as de Sitter.  Persistent lattice artifacts obstructing a smooth continuum limit of both Lorenztian and Euclidean theories, especially for interacting fields, are discussed in~\cite{Foster:2004yc, ern2004theory, brenner2008mathematical, Brower:2016moq}.
Note that both the receding mirror example and the de Sitter example are expanding cosmologies.  In each case the volume of spatial slices grows as a function of time, and so to maintain comparable spatial resolution we are required to increase the number of lattice sites as time advances.  

%and if we are to maintain the same UV cutoff on each Cauchy slice (i.e., our UV cutoff is defined as being a fixed multiple of the Planck length), then 

We now turn to the finite elements method, which provides a more sophisticated lattice discretization and  can handle the above circumstances in which  FDM fails.

\subsubsection{Finite elements method}
\label{Subsec:FEM1}

\begin{figure}
\begin{center}
\includegraphics[scale=.55]{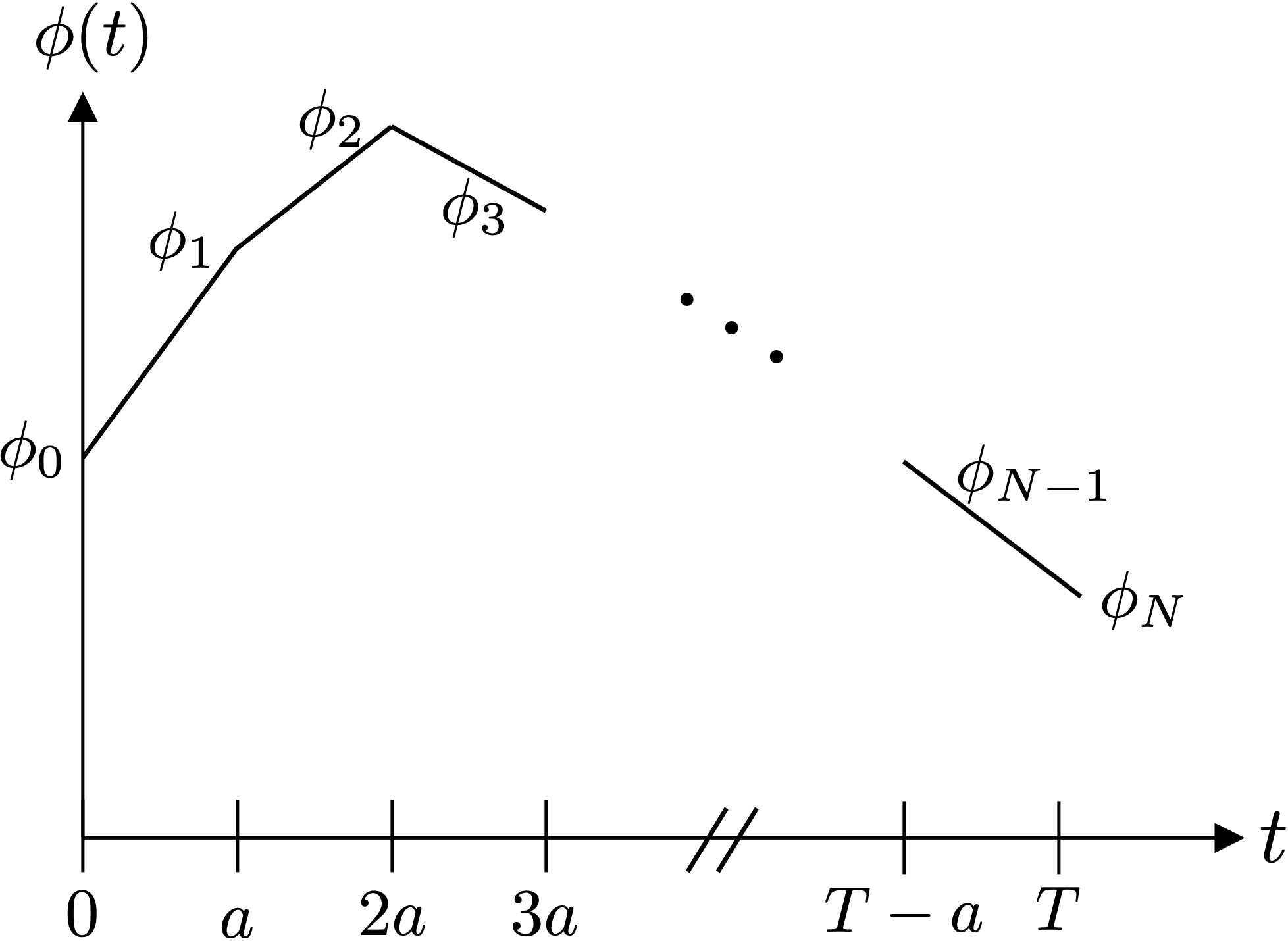}
\end{center}
\caption{Plot of a $\phi(t)$ in $\text{CPL}_a([0,T])$. \label{fig:philattice1}}
\end{figure}

Here we briefly review the FEM, which is widely used by engineers and other practitioners of PDE numerics~\cite{liu2021eighty}.
We first consider the 0+1 equations of a harmonic oscillator
\begin{equation}
\label{E:SHO1}
\ddot{\phi}(t) = - \phi(t)
\end{equation}
before turning to classical field theory in $d+1$ dimensions.
Now suppose we want to approximate $\phi(t)$ by a continuous, piecewise-linear function on $[0,T]$, where the linear segments have width $a = T/N$ playing  the role of our lattice cutoff.  We will call this space of functions $\text{CPL}_a([0,T])$.  See Figure~\ref{fig:philattice1} for a depiction of such a continuous piecewise function.  The most obvious way of expressing $\phi(t)$ is as
\begin{equation}
\label{E:manycases1}
\phi(t) = \begin{cases} \frac{\phi_1 - \phi_0}{a} \, t + \phi_0 &\text{if}\,\,\, 0 \leq t \leq a \\
\frac{\phi_2 - \phi_1}{a} \, (t - a) + \phi_0 &\text{if}\,\,\, a \leq t \leq 2a \\
\qquad \qquad \vdots & \qquad \quad \vdots \\
\frac{\phi_{j+1} - \phi_j}{a} \, (t - j a) + \phi_j &\text{if}\,\,\, j a \leq t \leq (j+1)a \\
\qquad \qquad \vdots & \qquad \quad \vdots \\
\frac{\phi_N - \phi_{N-1}}{a} \, (t - (T-a)) + \phi_{N-1} &\text{if}\,\,\, T - a \leq t \leq T
\end{cases}
\end{equation}
which is fully described by the values at the points $t = j a$ for $j = 0,1,...,N$, namely $\phi(j a) = \phi_j$. 

However, we have already run into a problem:  our equations of motion~\eqref{E:SHO1} cannot possibly be satisfied if $\phi(t)$ is continuous and piecewise linear.  In particular, $\dot{\phi}(t)$ is piecewise constant and discontinuous and thus $\ddot{\phi}(t)$ is a sum of delta distributions.  So the left and right-hand sides of~\eqref{E:SHO1} are not even in the same function space.

The standard way of addressing this problem is to weaken our criteria for what it means to solve~\eqref{E:SHO1} for $\phi \in \text{CPL}_a([0,T])$.  We say that a $\phi \in \text{CPL}_a([0,T])$ is a weak solution of~\eqref{E:SHO1} if
\begin{equation}
\int_0^T \! dt \, \psi(t) \big(\ddot{\phi}(t) + \phi(t)\big) = 0 \quad \text{for all }\psi \in \text{CPL}_a([0,T])\,.
\end{equation}
To avoid the appearance of delta distributions coming from the $\ddot{\phi}(t)$, we can integrate by parts to obtain
\begin{equation}
\label{E:FEMSHO1}
\int_0^T \! dt \, \big( - \dot{\psi}(t) \dot{\phi}(t) + \psi(t) \phi(t)\big) = 0 \quad \text{for all }\psi \in \text{CPL}_a([0,T])\,.
\end{equation}
Suppose a solution to this equation with Dirichlet boundary conditions $\phi(0) = \phi_i$ and $\phi(t) = \phi_f$ is $\widetilde{\phi}_a(t)$, where we have put in the subscript to emphasize the dependence on $a$.  What is the sense in which  $\widetilde{\phi}_a(t)$ achieves the correct continuum limit? 

Suppose $\widetilde{\phi}(t)$, here with no subscript, is an $L^2([0,T])$ solution to $\ddot{\phi} = - \phi$ with boundary Dirichlet boundary conditions $\phi(0) = \phi_i$ and $\phi(T) = \phi_f$.  Then it can be shown that
\begin{equation}
\lim_{a \to 0} \int_0^T \! dt \,\psi(t) \big(\widetilde{\phi}(t) - \widetilde{\phi}_a(t) \big) = 0 \quad \text{for all } \psi \in L^2([0,T])\,.
\end{equation}
This means that the our piecewise continuous solutions $\widetilde{\phi}_a(t)$ better and better approximate the $L^2([0,T])$ solution $\widetilde{\phi}(t)$ as we take $a \to 0$.  In fact, one can often prove stronger statements about convergence (e.g. strong convergence in $L^2$) but we will not need these results here.

While~\eqref{E:FEMSHO1} is a conceptually nice condition and illuminates the meaning of the continuum limit, it is not immediately obvious how to find a $\widetilde{\phi}_a(t)$ which satisfies~\eqref{E:FEMSHO1}.  Moreover it is not clear how to incorporate boundary conditions when finding such a solution.  Fortunately, there is an economical framework for finding weak solutions in the sense of~\eqref{E:FEMSHO1}, which elucidates the structure of piecewise continuous functions and allows for a more compact description than the right-hand side of~\eqref{E:manycases1}.

\begin{figure}
\begin{center}
\includegraphics[scale = .38]{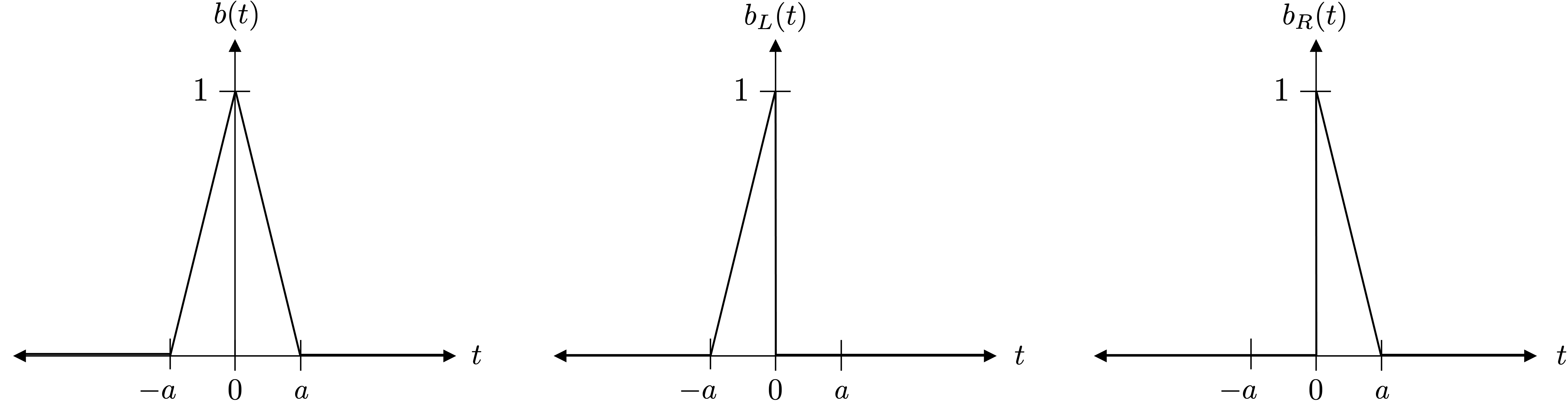}
\end{center}
\caption{Plots of $b(t)$, $b_L(t)$, and $b_L(t)$. \label{fig:basis1}}
\end{figure}

Let us begin by defining the so-called triangular basis functions.  We define the three functions $b(t)$, $b_L(t)$, and $b_R(t)$ by
\begin{align}
b(t) &= \begin{cases}
0 & \text{if} \,\,\, t < -a \\
\frac{t}{a}+1  & \text{if} \,\,\, - a \leq t \leq 0 \\
- \frac{t}{a} + 1 & \text{if} \,\,\, 0 \leq t \leq a \\
0 & \text{if}\,\,\,a < t 
\end{cases} \\ \nonumber \\
b_L(t) &= \begin{cases}
0 & \text{if} \,\,\, t < -a \\
\frac{t}{a}+1  & \text{if} \,\,\, - a \leq t \leq 0 \\
0 & \text{if}\,\,\,0 < t 
\end{cases} \\ \nonumber \\
b_R(t) &= \begin{cases}
0 & \text{if} \,\,\, t < 0 \\
- \frac{t}{a} + 1 & \text{if} \,\,\, 0 \leq t \leq a \\
0 & \text{if}\,\,\,a < t 
\end{cases}
\end{align}
These functions are piecewise linear, and are shown in Figure~\ref{fig:basis1}.
%We note for later than $b_L(t) + b_R(t) = b(t)$.
For our purposes, let
\begin{equation}
b_0(t) := b_R(t)\,, \qquad b_j(t) := b(t - j a)\,\,\,\text{for }\,j=1,...,N-1\,,\qquad b_N(t) := b_L(t - T)\,.
\end{equation}
Then we can rewrite our expression for $\phi(t)$ in~\eqref{E:manycases1} as
\begin{equation}
\phi(t) = \sum_{k = 0}^N \phi_k \, b_k(t)\,,
\end{equation}
and we will write a similar expansion $\psi(t) = \sum_{k=0}^N \psi_k \, b_k(t)$ for a test function $\psi(t)$.  Using these notations, we can express the weak solution condition in~\eqref{E:FEMSHO1} as
\begin{equation}
\label{E:FEMSHO2}
\sum_{i,j=1}^N  \psi_i  \left[\left(-\int_0^T \! dt\, \dot{b}_i(t) \,\dot{b}_j(t)\right)  + \left(\int_0^T \! dt\, b_i(t) \, b_j(t)\right) \right]\phi_j = 0 \quad \text{for all }\psi_0, \psi_1,...,\psi_N \in \mathbb{R}\,.
\end{equation}
Defining the stiffness matrix $S_{ij}$ and the mass matrix $M_{ij}$ by
\begin{align}
S_{ij} := -\int_0^T \! dt\, \dot{b}_i(t) \,\dot{b}_j(t)\,, \qquad M_{ij} := \int_0^T \! dt\, b_i(t) \, b_j(t)\,,
\end{align}
we find that~\eqref{E:FEMSHO2} is equivalent to
\begin{equation}
\label{E:FEMSHO3}
\sum_{j = 1}^N (S_{ij} + M_{ij}) \,\phi_j = 0 \quad \text{for }j=0,1,...,N\,.
\end{equation}
This is in fact a finite elements discretization of the continuum equation $\ddot{\phi} + \phi = 0$, with respect to our chosen basis functions $b_i(t)$.  Unraveling the definitions of $S_{ij}$ and $M_{ij}$, we find that the above is equal to
\begin{equation}
\label{E:FEMSHO4}
\frac{2 \phi_i - \phi_{i+1} - \phi_{i-1}}{a^2} + \frac{1}{3}\left( \phi_i + \frac{1}{2}(\phi_i + \phi_{i+1}) + \frac{1}{2}(\phi_i + \phi_{i+1})\right) = 0
\end{equation}
with some minor alternations near the boundaries at $i = 0$ and $i = N$ to account for edge effects.  We see that the first term is simply the lattice second derivative, and the second term approximates the value of $\phi(t)$ in the vicinity of $t = i a$.

Note that we could have obtained~\eqref{E:FEMSHO4} via the Euler-Lagrange equations of the lattice action
\begin{equation}
S_{\text{lattice}}[\{\phi_i\}] = - \frac{1}{2} \sum_{i,j = 1}^N \phi_i \,(S_{ij} + M_{ij})\, \phi_j
\end{equation}
which is the same as $\int_0^T dt \left(\frac{1}{2} \, \dot{\phi}^2 - \frac{1}{2}\, \phi^2\right)$ upon setting $\phi(t) = \sum_{k=0}^N \phi_k \, b_k(t)$.

We pause here to comment on what we have done so far, and summarize our existing procedure for lattice discretization via the FEM.  Suppose we have some differential equation for $\phi(t)$.  Then we carry out the following procedure:
\begin{enumerate}
\item Choose a suitable family of basis functions $b_i(t)$ whose span is the space $\text{CPL}_a([0,T])$.
\item Expand our function $\phi(t)$ in this basis as $\phi(t) = \sum_{k = 0}^N \phi_k \, b_k(t)$.
\item Plug this expansion of $\phi(t)$ into the equation for weak solutions of our differential equation, and plug in a similar expansion for the test function $\psi(t)$.  This reduces to a discrete system of equations which comprises the discretization of the continuum equations.  Alternatively, if we have a continuum action $S[\phi(t)]$ whose Euler-Lagrange equation are our differential equation, then we can simply let $\phi(t) = \sum_{k = 0}^N \phi_k \, b_k(t)$ to obtain a discrete action $S_{\text{lattice}}[\{\phi_i\}]$, and compute the new Euler-Lagrange equations which are our discretized equations of motion.
\end{enumerate}
This procedure generalizes to richer function spaces than $\text{CPL}_a([0,T])$, necessitating a more sophisticated choice of basis functions.  There is an enormous literature on this, but we will not need to delve into it here.

For us, the advantage of the FEM approach, besides its conceptual clarity, can be seen when we turn from 0+1 dimensions to 1+1 dimensions.  Before doing so, let us make a few structural observations about the 0+1 case to prepare for the 1+1 generalization.  We have organized the basis functions as $b_i(t)$ for $i=0,1,...,N$, where each basis function is associated to a single lattice site and the basis functions overlap.  There is a slightly different organization which fruitfully generalizes to the 1+1 setting.  Defining
\begin{equation}
b_{L,j}(t) := b_L(t - (j+1) a)\quad \text{for }\,\,j=0,...,N-1\,, \qquad b_{R,j}(t) := b_R(t - j a) \quad \text{for }\,\,j=0,...,N-1\,,
\end{equation}
we can associate $b_{L,j}(t)$ and $b_{R,j}(t)$ with the 1-dimensional edge connecting $t = j a$ to $t = (j+1)a$.  In this notation, we can write the piecewise version of $\phi(t)$ as
\begin{equation}
\phi(t) = \sum_{j=0}^{N-1} \phi_{j+1}\,b_{L,j}(t) + \sum_{j=0}^{N-1} \phi_{j}\,b_{R,j}(t)\,,
\end{equation}
where we have used the identity $b_L(t) + b_R(t) = b(t)$ to perform the above rewriting.  Indeed, $\{b_{L,j}(t),\, b_{R,j}(t)\}_{j=0}^{N-1}$ comprises a basis for $\text{CPL}_a([0,T])$, where we have two functions associated with each edge of the temporal lattice.

Next we examine the 1+1 massless Klein-Gordon equation, with action
$S[\phi(t)] = - \frac{1}{2} \int d^2 x \, \partial_\mu \phi \,\partial^\mu \phi$.
Although $\phi(t,x)$ naturally lives in $L^2(\mathbb{R}^2)$, we would like to approximate it by an appropriate family of continuous, piecewise linear functions on $\mathbb{R}^2$ (i.e., the $(t,x)$ plane).  To specify this family, we can consider various lattice discretizations of $\mathbb{R}^2$.  The most obvious is a rectangular grid with lattice spacings $\delta t, \delta x$.  For this, we can construct a basis $\{b_{i,j}(t,x)\}_{i,j \in \mathbb{Z}}$ where
\begin{equation}
b_{i,j}(t,x) := b( \frac{a}{\delta t} (t - i \,\delta t)) \,b( \frac{a}{\delta x}(x - j \,\delta x))
\end{equation}
is associated to the $i,j$ vertex.  Equivalently, we can use the basis $\{b_{i,j}^{(1)}(t,x),\,b_{i,j}^{(2)}(t,x), b_{i,j}^{(3)}(t,x), b_{i,j}^{(4)}(t,x)\}_{i,j \in \mathbb{Z}}$ where
\begin{align}
b_{i,j}^{(1)}(t,x) &= b_L(\frac{a}{\delta t}(t - (i+1)\delta t)) \,b_L(\frac{a}{\delta x}(x - (j+1)\delta x))\,, \,\, b_{i,j}^{(2)}(t,x) = b_L(\frac{a}{\delta t}(t - (i+1)\delta t) \,b_R(\frac{a}{\delta x}(x - j \,\delta x))\,,\nonumber \\
b_{i,j}^{(3)}(t,x) &= b_R(\frac{a}{\delta t}(t - i \,\delta t))\,b_L(\frac{a}{\delta x}(x - (j+1) \delta x))\,,\,\, b_{i,j}^{(4)}(t,x) = b_R(\frac{a}{\delta t}(t - i \,\delta t)) \,b_R(\frac{a}{\delta x}(x - j \,\delta x))\,.
\end{align}
Here, each quadruplet of basis elements $b_{i,j}^{(1)}(t,x),\,b_{i,j}^{(2)}(t,x), b_{i,j}^{(3)}(t,x), b_{i,j}^{(4)}(t,x)$ is associated with a single plaquette whose lower left vertex is the $i,j$ vertex.

Instead of considering a rectangular lattice, we will opt for the most flexible lattice of all: a triangular lattice.  Indeed, any planar lattice can be refined into a triangular lattice by adding more edges.  We emphasize that there is enormous freedom in choosing a triangulation; for hyperbolic PDE's such as wave equations, we should be careful that our triangulation is in compliance with the CFL condition~\cite{courant1967partial} which affords numerical stability of the lattice approximation as we take the continuum limit.\footnote{In words, the CFL condition for massless free field theory is that the past lightcone of any point on the lattice must be inside the lattice domain of dependence of that point (i.e.~the domain of dependence induced by the lattice equations of motion).}  We will express our lattice $\mathscr{L} = \{\Delta_i\}$ as a collection of triangular plaquettes $\Delta_i$\,, and to each plaquette $\Delta_i$ we associate an ordered list of its vertices in $\mathbb{R}^2$, namely $(v_{i,1}, v_{i,2}, v_{i,3}) = \big((x_{i,1}, y_{i,1}), (x_{i,2}, y_{i,2}), (x_{i,3}, y_{i,3})\big)$.  Next we define our family of basis functions for $\mathscr{L}$.  Let $\chi_{\Delta}(t,x)$ be the characteristic function for the interior of the `standard' triangle $\Delta$ in $\mathbb{R}^2$ with vertices $(0,0)$, $(1,0)$ and $(0,1)$.  We can write $\chi_{\Delta}(t,x)$ explicitly in terms of a product of Heaviside step functions as $\chi_{\Delta}(t,x) = \theta(1-t-x) \,\theta(t) \,\theta(x)$.  Then we define the three functions
\begin{equation}
B_{\Delta}^{(1)}(t,x) := (1 - t - x)\,\chi_{\Delta}(t,x)\,, \quad B_{\Delta}^{(2)}(t,x) := t\,\chi_{\Delta}(t,x)\,,\quad  B_{\Delta}^{(3)}(t,x) := x\,\chi_{\Delta}(t,x)\,.
\end{equation}
We can map the `standard' triangle $\Delta$ to any triangle in $\mathbb{R}^2$ via an affine transformation.  In particular, if we want to map $\big((0,0), (1,0), (0,1)\big)$ to $\big((x_1, y_1), (x_2, y_2), (x_3, y_3)\big)$, then we can leverage the affine transformation
\begin{equation}
A : \begin{bmatrix}
x \\ y
\end{bmatrix} \longmapsto \begin{bmatrix}
-x_1 + x_2  & -x_1 + x_3 \\ -y_1 + y_2 & - y_1 + y_3
\end{bmatrix} \begin{bmatrix}
x \\ y
\end{bmatrix} + \begin{bmatrix}
x_1 \\ y_1
\end{bmatrix}
\end{equation} 
and transform the $B_{\Delta}^{(j)}(t,x)$ functions to $B_{\Delta}^{(j)}(A^{-1}(t,x))$  accordingly.  We should be careful to map between triangles with the same orientation in the plane.  Let $B_{\Delta_i}^{(1)}(t,x), B_{\Delta_i}^{(2)}(t,x), B_{\Delta_i}^{(3)}(t,x)$ correspond to the basis functions above, affinely transformed so that their vertices match the vertices of the plaquette $\Delta_i$.  Then we take our family of basis functions to be
\begin{equation}
\left\{B_{\Delta_i}^{(1)}(t,x), B_{\Delta_i}^{(2)}(t,x), B_{\Delta_i}^{(3)}(t,x)\right\}_{\Delta_i \in \mathscr{L}}
\end{equation}
and denote its real-valued span by the family of functions $\text{CPL}_{\mathscr{L}}(\mathbb{R}^2)$.  We can accordingly expand $\phi(t,x)$ in this basis as
\begin{equation}
\phi(t,x) = \sum_{i} \left(\phi_{v_{i,1}} \, B_{\Delta_i}^{(1)}(t,x) + \phi_{v_{i,2}} \, B_{\Delta_i}^{(2)}(t,x)  + \phi_{v_{i,3}} \, B_{\Delta_i}^{(3)}(t,x) \right) \,.
\end{equation}
Plugging this into the continuum free action, we obtain a lattice action
\begin{equation}
\label{E:FEMfree1}
S_{\text{lattice}}[\{\phi_v\}] = - \frac{1}{2}\sum_{v,v' \in V(\mathscr{L})} \phi_v \, Q_{v,v'} \, \phi_{v'}
\end{equation}
for a corresponding stiffness matrix $Q_{v,v'}$\,, where $V(\mathscr{L})$ is the set of vertices of $\mathscr{L}$.  We compute some explicit examples below.

\begin{figure}
\begin{center}
\includegraphics[scale=.5]{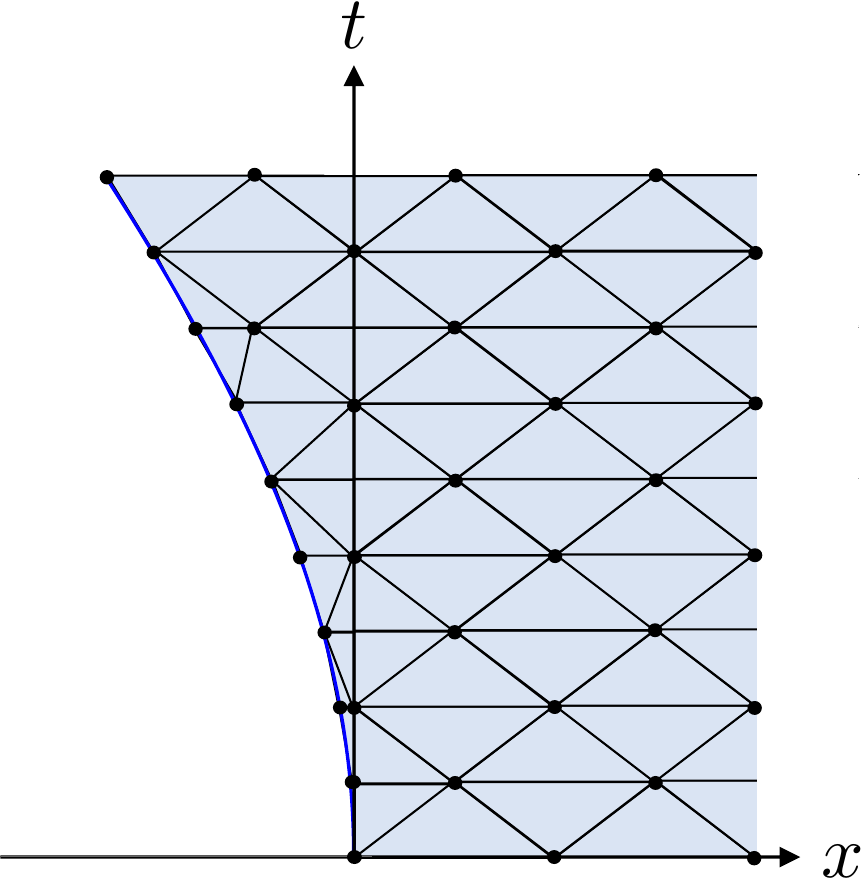}
\end{center}
\caption{A triangular FEM lattice regularization of a 1+1 scalar field in a spacetime with a moving boundary.  The moving boundary is piecewise-linear-approximated by the lattice discretization, which is continuous for any cutoff scale; this is in contrast to the FEM saw-tooth approximation of Figure~\ref{fig:timedepbdy2}.  Note that the number of lattice sites on each Cauchy slice is increasing with time. \label{fig:timedepbdy3}}
\end{figure}

Having triangular plaquettes affords us big advantages over rectangular plaquettes.  The first is that we can accommodate arbitrarily-shaped spacetime boundaries without making a saw-tooth approximation.  That is, we can approximate spacetime boundaries in a piecewise-linear fashion that appropriately converge to smooth boundaries in the continuum limit, i.e.~as we consider a sequence of progressively finer triangulations.  This will preclude the divergence of the energy flux from the moving boundary which was a pathology of the saw-tooth approximation.  See Figure~\ref{fig:timedepbdy3} for the example of a free scalar with a moving left boundary, and notice that the triangular lattice both well-approximates the moving boundary and maintains being continuous in the continuum limit.  The second advantage is that we have a more flexible framework for coupling our fields to a curved background, although we will not explore this presently.  The third advantage is that we can more readily incorporate a changing number of lattice sites on spacelike Cauchy slices, i.e.~when the spatial volume is changing as a function of time.

\subsection{Nonunitary isometric quantum evolution }
\label{Subsec:pathintegral1}

So far we have explored lattice discretizations of classical field equations which converge to desired continuum equations in an appropriate limit.  For situations with curved spacetime boundaries or geometrically non-trivial interiors, e.g.~for expanding cosmologies, we have seen that the FEM provides an advantageous lattice discretization at the classical level.  But what about the quantum theory?  The most natural way to proceed is to simply perform a path integral quantization of the FEM lattice theory, e.g.~\eqref{E:FEMfree1}.  Our main focus will be to understand how the Lorentzian time evolution is instantiated by the path integral in a quantum manner, in particular since the number of lattice sites on a Cauchy slice will be growing with time for our examples of interest.  There has been related work using canonical quantization in~\cite{Foster:2004yc, Dittrich:2013jaa, Hohn:2014uvt, Hohn:2014rba}.

We remark that in the Euclidean setting, the FEM has been recently leveraged to great effect for lattice simulations of quantum field theories on curved backgrounds~\cite{Brower:2016vsl, Brower:2018szu, Brower:2019kyh}; see~\cite{Brower:2016moq} for an overview.  The formalism of these papers in fact goes beyond the usual FEM, and incorporates refinements custom-tailored for quantum field theories.  While similar modifications will certainly be useful in our Lorentzian setting, here we will consider examples that are simple enough that modifications beyond the FEM are unnecessary.

\begin{figure}
\begin{center}
\includegraphics[scale=.55]{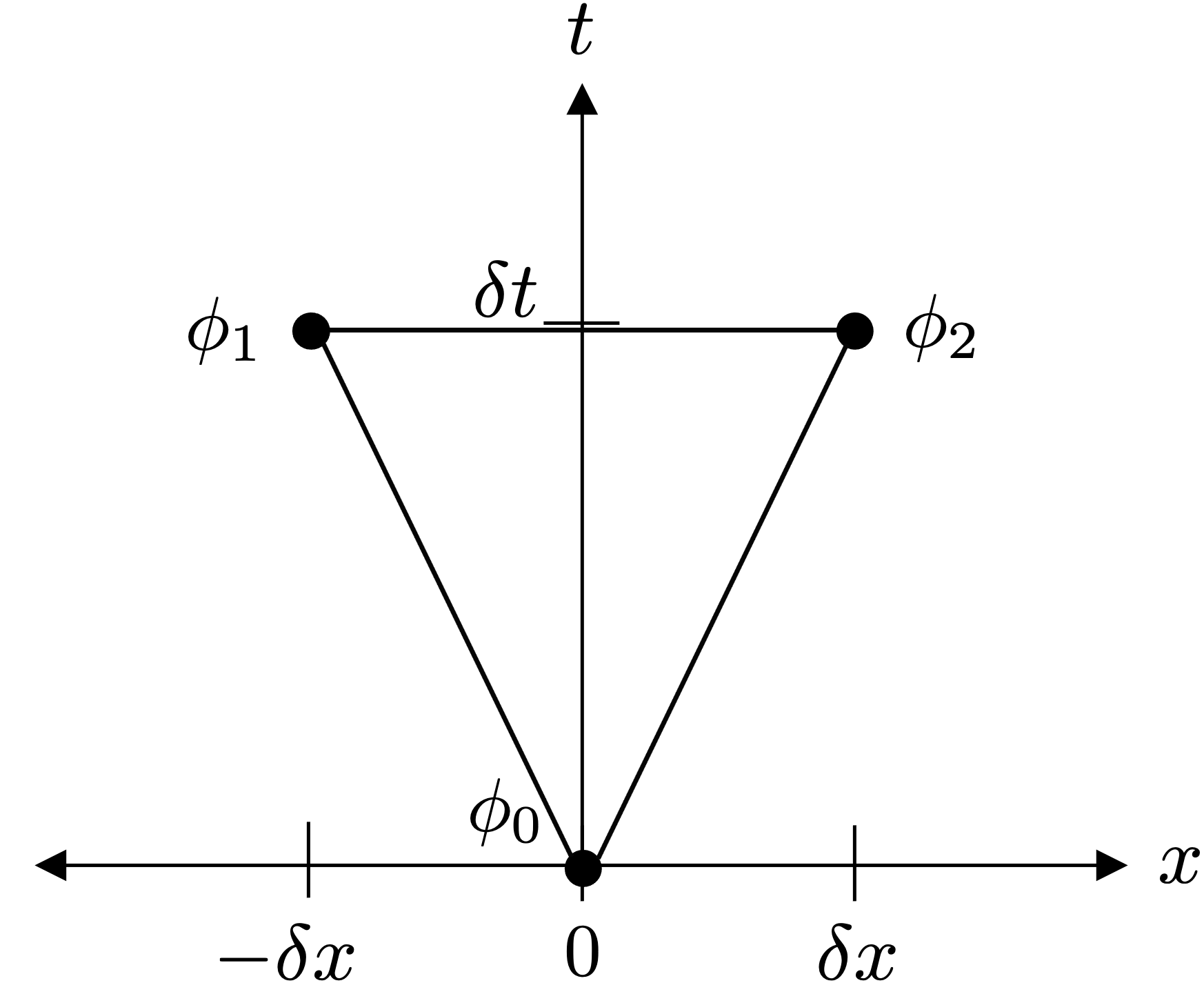}
\end{center}
\caption{The triangle $\widetilde{\Delta}$ in $\mathbb{R}^2$ with vertices $(t,x) = (0,0)$, $(\delta t,\, -\delta x)$ and $(\delta t,\,\delta x)$.  We have labeled the vertices by their corresponding field values $\phi_0$, $\phi_1$, $\phi_2$. \label{fig:triangle}}
\end{figure}

Let us begin with an illuminating a toy example, namely a field theory on a triangle with three vertices as seen in Figure~\ref{fig:triangle}.  We call the triangle $\widetilde{\Delta}$, with vertices $(t,x) = (0,0)$, $(\delta t,\, -\delta x)$ and $(\delta t,\,\delta x)$.  In the case of a free massless scalar field, we can expand $\phi(t,x)$ as
\begin{equation}
\label{E:phitriangleexpansion1}
\phi(t,x) = \phi_0 \, B_{\widetilde{\Delta}}^{(0)}(t,x) + \phi_1 \, B_{\widetilde{\Delta}}^{(1)}(t,x) + \phi_2 \, B_{\widetilde{\Delta}}^{(2)}(t,x) 
\end{equation}
and then plug this into $S[\phi(t,x)] = - \frac{1}{2} \int d^2 x \, \partial_\mu \phi \,\partial^\mu \phi$ to obtain a lattice action.
%\begin{equation}
%\label{E:singletriangle1}
%S_{\text{lattice}}[\{\phi_0, \phi_1, \phi_2\}] = - \frac{1}{4} \, \phi_0^2 + \frac{3}{16} (\phi_1^2 + \phi_2^2) + \frac{1}{4} \, \phi_0 (\phi_1 + \phi_2) - \frac{5}{8} \, \phi_1 \phi_2\,.
%\end{equation}
%where the values of the parameters are
%\begin{equation}
%\alpha = \frac{1}{4}\,, \qquad \beta = \frac{3}{16}\,, \qquad \gamma = \frac{1}{4}\,, \qquad \delta = \frac{5}{8}\,.
%\end{equation}
%While these exact values will not be so important for us, we note that the $\phi_0^2$ being the `wrong sign' is an artifact of having a single spatial lattice site at time $t = 0$. 
Let us see what happens when we quantize this lattice theory.

It will be essential to leverage an $i\varepsilon$ prescription that will enable our path integral to converge, and which will contribute other essential ingredients to our analysis.  Accordingly, we augment our continuum action to be
\begin{equation}
S[\phi(t,x)] = - \frac{1}{2}\int d^2 x \, \eta^{\mu \nu} \partial_\mu \phi \, \partial_\nu \phi  + i \varepsilon \cdot \frac{1}{2}\int d^2 x \,  \delta^{\mu \nu} \partial_\mu \phi \, \partial_\nu \phi\,  \,.
\end{equation}
Plugging in~\eqref{E:phitriangleexpansion1}, the corresponding lattice action is
\begin{equation}
\label{E:singletriangle2}
S_{\text{lattice}}[\{\phi_0, \phi_1, \phi_2\}] = \frac{\delta x}{\delta t}\, \cdot \frac{1 + i \varepsilon}{8}\,\left(2 \phi_0 - (\phi_1 + \phi_2)\right)^2 - \frac{\delta t}{\delta x}\cdot \frac{1 - i \varepsilon}{8}\,(\phi_1-\phi_2)^2
\end{equation}
% Here we have simply implemented the usual $i \varepsilon$ prescription at the level of the continuum action, and then plugged in~\eqref{E:phitriangleexpansion1} as usual.
which has the property that
\begin{equation}
\text{Re}\left\{i \, \frac{\delta^2 S_{\text{lattice}}}{\delta \phi_i \delta \phi_j}\right\} \preccurlyeq 0\,,
\end{equation}
namely the real part of the second derivative of the action times $i$ is negative semi-definite so that our path integral may converge.
%We will set $\delta t = 1$ and $\delta x = 2$ to be compliant with the CFL condition $\frac{\delta t}{\delta x} \leq 1$, although in the present example this is not strictly necessary since there are only three lattice sites.
Our desired path integral is simply
\begin{equation}
C \int d\phi_0 \, d\phi_1 \, d\phi_2\, e^{i \, S_{\text{lattice}}[\{\phi_0, \phi_1, \phi_2\}]}
\end{equation}
with the action as in~\eqref{E:singletriangle2}, for some overall constant $C$ to be determined.

An immediate question arises: is our path integral unitary?  In a Schr\"{o}dinger formalism, our system at time $t = 0$ is described by a wavefunction $\Psi(\phi_0)$ in $\mathcal{H}_0$, and our system at time $t = 1$ is described by a wavefunction $\Psi(\phi_1, \phi_2)$ in $\mathcal{H}_1 \otimes \mathcal{H}_2$.  Thus our forward time evolution must map $\mathcal{H}_0 \to \mathcal{H}_1 \otimes \mathcal{H}_2$ and our backward time evolution must map $\mathcal{H}_1 \otimes \mathcal{H}_2 \to \mathcal{H}_0$.  Let us see how this manifests at the level of the path integral.  The propagator from $t = 0$ to $t = 1$ is just
\begin{equation}
\label{E:kernel1}
K( \phi_1, \phi_2\,;\,\phi_0) = C \, e^{i \, S_{\text{lattice}}[\{\phi_0, \phi_1, \phi_2\}]}\,.
\end{equation}
What happens if we evolve from $t = 0$ to $t = 1$, and then back to $t = 0$?  This is expressed by
\begin{equation}
\int d\phi_1 \, d\phi_2  \, K^*(\phi_1, \phi_2\,;\,\phi_0') \, K(\phi_1, \phi_2\,;\,\phi_0)= |C|^2 \, \frac{2\pi}{\varepsilon}\, \exp\left(- \frac{\delta x}{\delta t}\,\frac{1 + \varepsilon^2}{4\,\varepsilon} \, (\phi_0 - \phi_0')^2 \right)\,.
\end{equation}
As $\varepsilon \to 0$, the above tends to $|C|^2\,4\sqrt{\frac{\delta t}{\delta x} \frac{\pi^3}{\varepsilon}}\, \delta(\phi_0 - \phi_0')$, and so we should set
\begin{equation}
\label{E:const1}
C = \frac{1}{2}\left(\frac{\delta x}{\delta t}\,\frac{\varepsilon}{\pi^3}\right)^{1/4}
\end{equation}
to obtain $\delta(\phi_0 - \phi_0')$.  This condition can be expressed more algebraically.  Let $\mathcal{K}$ be the operator form of the kernel, i.e.~with matrix elements $\langle \phi_1, \phi_2 |\,\mathcal{K}\,|\phi_0\rangle = K( \phi_1, \phi_2\,;\,\phi_0)$.  We have $\mathcal{K} : \mathcal{H}_0 \to \mathcal{H}_1 \otimes \mathcal{H}_2$,
%where $\mathcal{H}_i$ is the Hilbert space associated with site $i$, 
and the above calculation tells us that
\begin{equation}
\label{E:isometry1}
\mathcal{K}^\dagger \mathcal{K} = \mathds{1}_{\mathcal{H}_0}\,.
\end{equation}
That is, if we evolve from $t = 0$ to $t = 1$ and then back, we find the identity operator on the $t = 0$ Hilbert space $\mathcal{H}_0$.

The condition~\eqref{E:isometry1} tells us that $\mathcal{K}$ is a Hilbert space isometry.  That is, if we take any two states $|\Phi\rangle$ and $|\Phi'\rangle$ in $\mathcal{H}_0$ at the initial time $t = 0$ and evolve them via $\mathcal{K}$, then their inner product is preserved:
\begin{equation}
\label{E:isometry2}
\langle \Phi' | \mathcal{K}^\dagger \mathcal{K} |\Phi\rangle = \langle \Phi'| \Phi\rangle\,.
\end{equation}
Note that $\mathcal{K}|\Phi\rangle$ and $\mathcal{K} |\Phi'\rangle$ are each states in $\mathcal{H}_1 \otimes \mathcal{H}_2$ at time $t = 1$.  We also have that $\mathcal{K}^\dagger : \mathcal{H}_1 \otimes \mathcal{H}_2 \to \mathcal{H}_0$; in fact, the condition~\eqref{E:isometry1} constrains the possibilities for $\mathcal{K}\mathcal{K}^\dagger$ which is a map from $\mathcal{K}^\dagger : \mathcal{H}_1 \otimes \mathcal{H}_2 \to \mathcal{H}_0$.  In particular,
\begin{equation}
(\mathcal{K} \mathcal{K}^\dagger) (\mathcal{K} \mathcal{K}^\dagger) = \mathcal{K} \mathcal{K}^\dagger\,,
\end{equation}
i.e.~$\mathcal{K} \mathcal{K}^\dagger$ is idempotent, and since it is also Hermitian we conclude that $\mathcal{K} \mathcal{K}^\dagger$ is a projector.  Moreover, the rank of the projector is readily seen to be the same as the rank of $\mathcal{K}^\dagger \mathcal{K} = \mathds{1}_{\mathcal{H}_0}$, but since the rank of $\mathds{1}_{\mathcal{H}_0}$ is infinite we do not learn much about which projector we have.  If we had $\mathcal{K} \mathcal{K}^\dagger = \mathds{1}_{\mathcal{H}_1 \otimes \mathcal{H}_2}$, then $\mathcal{K}$ would be unitary.  However, this turns out not to be the case: in Appendix~\ref{App:isom1} we show that $\mathcal{K} \mathcal{K}^\dagger$ has a non-trivial nullspace,
%\footnote{See Appendix~\ref{App:spaces} for a discussion of some subtleties about the definition of the nullspace.}
which definitively establishes that $\mathcal{K}$ is not unitary and is instead merely an isometry.

% This is a possibility since $L^2(\mathbb{R})$ is isomorphic to $L^2(\mathbb{R}) \otimes L^2(\mathbb{R})$.  But more generally we might have $\mathcal{K} \mathcal{K}^\dagger \not = \mathds{1}_{L^2(\mathbb{R}) \otimes L^2(\mathbb{R})}$, in which case $\mathcal{K}$ is merely an isometry and \textit{not} a unitary.  We will shortly check which of these cases holds true.

In our examples, subtleties arose because $\mathcal{H}_0, \mathcal{H}_1, \mathcal{H}_2$ are infinite-dimensional; for some details of these function spaces and how they interface with the $i\varepsilon$ prescription, see Appendix~\ref{App:spaces}.  We briefly pause to remark that if $\mathcal{K}$ was a map between finite-dimensional Hilbert spaces, the non-unitarity of $\mathcal{K}$ would be immediate.  For instance, suppose $\dim(\mathcal{H}_0) < \dim(\mathcal{H}_1) = \dim(\mathcal{H}_2)< \infty$.  If $\mathcal{K}^\dagger \mathcal{K} = \mathds{1}_{\mathcal{H}_0}$, then the rank of $\mathcal{K} \mathcal{K}^\dagger$ must be $\dim(\mathcal{H}_0)$, and so $\mathcal{K} \mathcal{K}^\dagger \not = \mathds{1}_{\mathcal{H}_1 \otimes \mathcal{H}_2}$ since the right-hand side has rank $\dim(\mathcal{H}_1)$.  Thus $\mathcal{K}$ would merely be an isometry and \textit{not} a unitary.  In section~\ref{Sec:embedding}, we will consider tensor networks which provide toy models of de Sitter; here we will indeed have isometries between finite-dimensional Hilbert spaces as our time evolution.

%For instance, consider $\Psi(\chi_1',\chi_2) = $
%The net result is that $\mathcal{K}\mathcal{K}^\dagger$ projects $\Psi(\phi_1, \phi_2)$ onto the subspace~\textcolor{red}{[Is this actually a subspace since the wavefunctions are not normalizable in $L^2(\mathbb{R}) \otimes L^2(\mathbb{R})$]}
%\begin{equation}
%S = \text{span}\{\}
%\end{equation}
%and so we write
%\begin{equation}
%\end{equation}

%\mathcal{K} \mathcal{K}^\dagger = \mathcal{P}_S
%\end{equation}
%where $\mathcal{P}_S$ is the projector onto $S$; notably, this is not the identity $\mathds{1}_{L^2(\mathbb{R})\otimes L^2(\mathbb{R})}$.  We thus see that $\mathcal{K}$ is an isometry and not a unitary.

Having worked through our toy example, we can now graduate to a full-fledged field theory with an arbitrary number of lattice sites.  Suppose we choose a triangular lattice $\mathscr{L}$ of $\mathbb{R}^2$ (interpreted as the $(t,x)$ plane), and we consider the propagator between two adjacent Cauchy slices.  Let the field variables on the `past' Cauchy slice be denoted by the vector $\vec{\phi}_P$ with $|P|$ components, and the field variables on the `future' Cauchy slice by $\vec{\phi}_F$ with $|F|$ components.  We will suppose that $|F| > |P|$, namely that there are a greater number of lattice points on the future Cauchy slice than on the past Cauchy slice.  Then for a free scalar field theory, our propagator between these slices will have the form
\begin{equation}
\label{E:Kprop1}
K(\vec{\phi}_F\,;\,\vec{\phi}_P) = C\,e^{i \left(\vec{\phi}_P \cdot Q \cdot \vec{\phi}_P + \vec{\phi}_F \cdot R \cdot \vec{\phi}_P + \vec{\phi}_F \cdot S \cdot \vec{\phi}_F \right) - \varepsilon \, T(\vec{\phi}_P, \vec{\phi}_F)}
\end{equation}
where $T(\vec{\phi}_P, \vec{\phi}_F)$ is quadratic in the $\vec{\phi}_P$'s and $\vec{\phi}_F$'s, and moreover is positive definite.  Let us compute the kernel which propagates us from past to future to past.  This is
\begin{equation}
\int d\vec{\phi}_F \, K^*(\vec{\phi}_F\,;\,\vec{\phi}_P')  \,  K(\vec{\phi}_F\,;\,\vec{\phi}_P) = |C|^2\,e^{i \left(\vec{\phi}_P \cdot Q \cdot \vec{\phi}_P - \vec{\phi}_P' \cdot Q \cdot \vec{\phi}_P' + \vec{\phi}_F \cdot R \cdot (\vec{\phi}_P - \vec{\phi}_P') \right) - \varepsilon \, T(\vec{\phi}_P, \vec{\phi}_F) - \varepsilon \, T(\vec{\phi}_P', \vec{\phi}_F)}\,.
\end{equation}
Noticing the $e^{i(\vec{\phi}_F \cdot R \cdot (\vec{\phi}_P - \vec{\phi}_P'))}$ factor, it seems that if we performed the integral then in the $\varepsilon \to 0$ limit we would obtain $\sim \frac{1}{\varepsilon^{|F|/2}} \, \delta^{|F|}\!\big( R\cdot(\vec{\phi}_P - \vec{\phi}_P')  \big)$.  But this is too quick; we need to take into account that $R$ is a $|F| \times |P|$ matrix for $|F| > |P|$.  Since the rank of $R$ is actually $|P|$, in the $\varepsilon \to 0$ limit we in fact have
\begin{equation}
|C|^2 \, \frac{\#}{\varepsilon^{\frac{|F|-|P|}{2}}} \, \delta^{|P|}\!\big( \vec{\phi}_P - \vec{\phi}_P' \big)
\end{equation}
where $\#$ is a constant.  Letting $C = \varepsilon^{\frac{|F|-|P|}{4}}/\#^{1/2}$, we are left with
\begin{equation}
\delta^{|P|}\!\big( \vec{\phi}_P - \vec{\phi}_P' \big)\,,
\end{equation}
and so $\mathcal{K}^\dagger \mathcal{K}$ is the identity.  Hence $\mathcal{K}$ is a Hilbert space isometry.  Similarly, if we examine future to past to future propagation we find
\begin{equation}
\int d\vec{\phi}_P \,  K(\vec{\phi}_F\,;\,\vec{\phi}_P)\,K^*(\vec{\phi}_F'\,;\,\vec{\phi}_P)  = |C|^2\,e^{i \left(\vec{\phi}_F \cdot S \cdot \vec{\phi}_F - \vec{\phi}_F' \cdot S \cdot \vec{\phi}_F' + (\vec{\phi}_F - \vec{\phi}_F')\cdot R \cdot \vec{\phi}_P \right) - \varepsilon \, T(\vec{\phi}_P, \vec{\phi}_F) - \varepsilon \, T(\vec{\phi}_P, \vec{\phi}_F')}\,.
\end{equation}
Considering the $e^{i  (\vec{\phi}_F - \vec{\phi}_F')\cdot R \cdot \vec{\phi}_P }$ factor, we realize that since $R$ has rank $|P|$, the integral cannot possibly equal $\delta^{|F|}(\vec{\phi}_F - \vec{\phi}_F')$, and so $\mathcal{K} \mathcal{K}^\dagger$ is not the identity.  We again conclude that $\mathcal{K}$ is not unitary and is instead just an isometry.  

In an expanding spacetime, we have a sequence of Cauchy slices such that the number of lattice sites per Cauchy slice is increasing with time.  If $\mathcal{K}_{2 \leftarrow 1}$ is the propagator from the first to second Cauchy slice, $\mathcal{K}_{3 \leftarrow 2}$ from the second to third, and so on, then our analysis above establishes that each $\mathcal{K}_{i+1 \leftarrow i}$ is a Hilbert space isometry.  Then it is readily checked that for $i < j$,
\begin{equation}
\mathcal{K}_{j \leftarrow i} := \mathcal{K}_{j \leftarrow j-1} \cdots \mathcal{K}_{i+2 \leftarrow i+1} \mathcal{K}_{i+1 \leftarrow i}
\end{equation}
is also Hilbert space isometry, since $\mathcal{K}_{i \to j}^\dagger \mathcal{K}_{i \to j}$ is the identity on the Hilbert space corresponding to the $i$th Cauchy slice, and $\mathcal{K}_{i \to j} \mathcal{K}_{i \to j}^\dagger$ is a non-identity projection.

Although we have focused on free field theory here, in Appendix~\ref{App:isom2} we discuss interacting theories.

Let us summarize.  First, we argued that even at the classical level, performing a lattice discretization of a field theory can be delicate in the presence of time-dependent backgrounds.  To have a lattice discretization which obtains the appropriate continuum limit, we often need to resort to the FEM which is vastly more flexible than the FDM (i.e., the current standard practice in lattice field theory).  The FEM accommodates moving boundaries and expanding cosmologies, all while giving us the flexibility to have different numbers of lattice sites on each Cauchy slice.  When we quantized Lorentzian FEM lattice actions via the path integral, we found that time evolution was not generally unitary.  Rather, in an expanding background we found that the path integral renders time evolution into a Hilbert space isometry.

\subsection{The moving mirror}
\begin{figure}[t]
\begin{center}
\includegraphics[scale=.5]{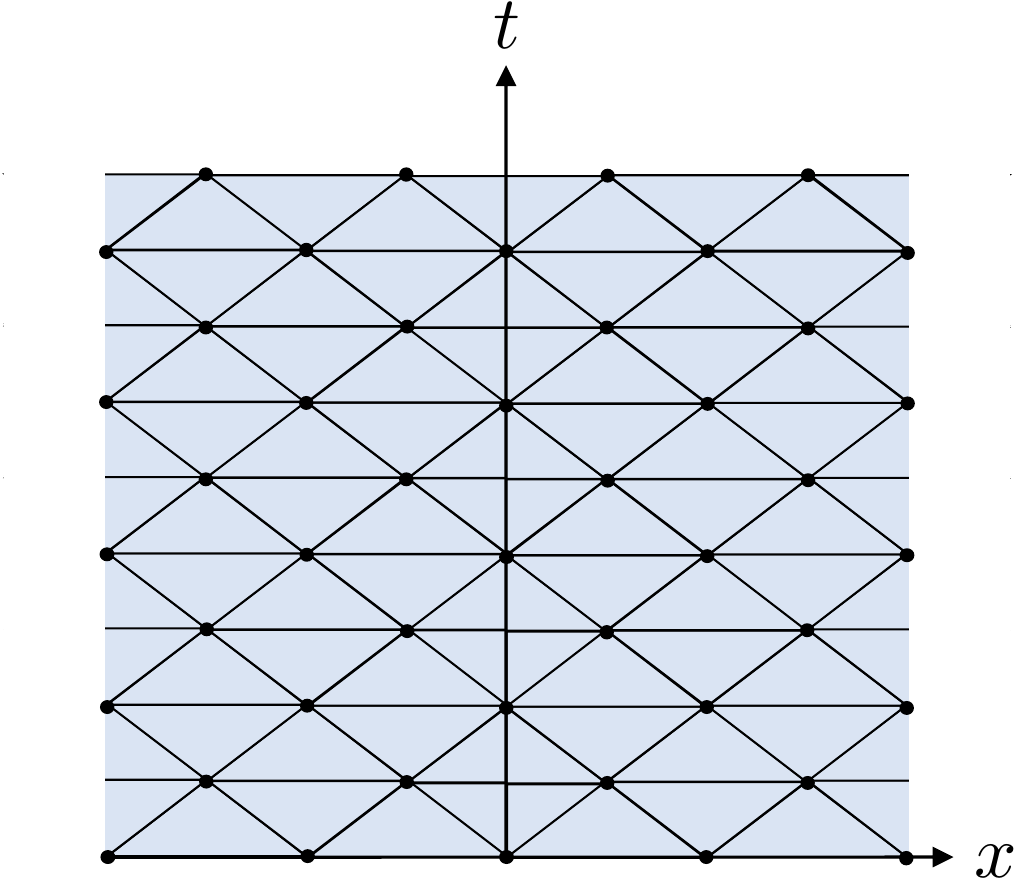}
\end{center}
\caption{A triangulation of $\mathbb{R}^2$ for $t \geq 0$, built out of isosceles triangles; the vertices of the triangles are shown as bolded dots.  Note that the $t$-axis passes over triangles, but does not not divide them into smaller triangles.\label{Fig:newtriangulation1}}
\end{figure}

Here we construct a finite elements discretization of the uniformly accelerating mirror in 1+1 dimensions.  There are many options for triangulation schemes, and we choose a particularly simple one.
% as a toy model of an expanding universe.  %1+1 de Sitter can be treated in a similar fashion, e.g.~by adapting the Euclidean tools from~\ref{Brower:2019kyh}, we do not do so here since the moving mirror will be illustrative enough.  We conclude the section by discussing more general features of a universe where unitary evolution has been replaced by isometric evolution as the universe expands.
We begin by consider a 1+1 free massless scalar field for time $t \geq 0$, at first without a mirror boundary.  Let us leverage the triangulation shown in Figure~\ref{Fig:newtriangulation1}; here the triangles are all isosceles and have width $2 \,\delta x$ and height $\delta t$.  In order to satisfy the CFL condition for numerical stability, we choose $\frac{\delta t}{\delta x} < 1$.  In Figure~\ref{Fig:newtriangulation1}, we label vertices according to
\begin{align}
\label{E:sidebyside}
    \includegraphics[scale=.45, valign = c]{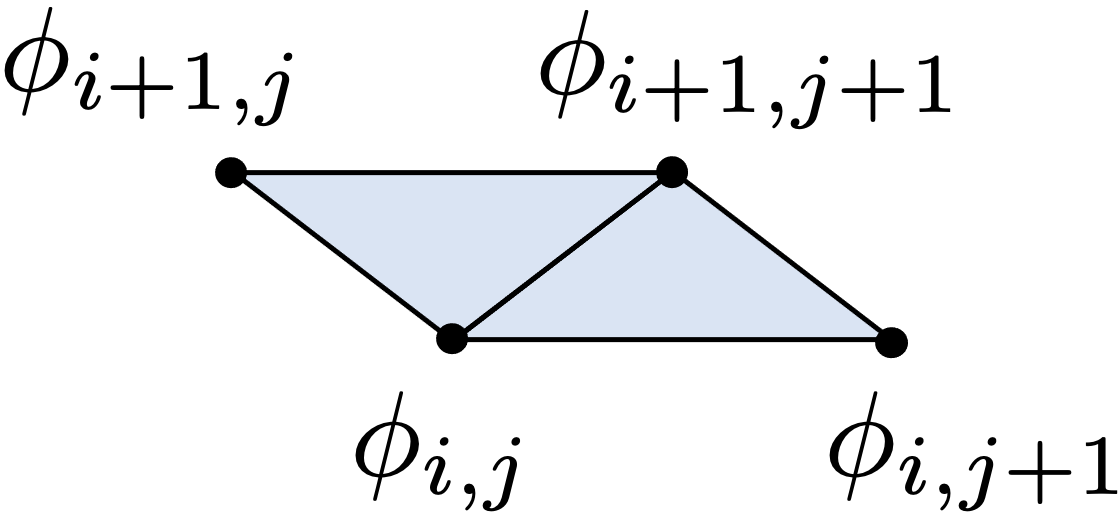}
\end{align}
which will be useful for the FEM machinery which follows.  Let $\bigtriangledown_{i,j}$ denote a triangular plaquette with its bottom vertex labeled by $\phi_{i,j}$, its upper left vertex labeled by $\phi_{i+1,j}$, and its upper right vertex labeled by $\phi_{i+1,j+1}$.  Similarly, let $\bigtriangleup_{i,j}$ denote a plaquette with its bottom left vertex labeled by $\phi_{i,j}$, its upper vertex labeled by $\phi_{i+1,j+1}$, and its lower right vertex labeled by $\phi_{i,j+1}$.  Indeed,~\eqref{E:sidebyside} shows a $\bigtriangledown_{i,j}$ and a $\bigtriangleup_{i,j}$ side-by-side.  Each of these plaquettes contributes to the total lattice action as
\begin{align}
S[\bigtriangledown_{i,j}] &= \frac{\delta x}{\delta t}\, \cdot \frac{1 + i \varepsilon}{8}\,\left(2 \phi_{i,j} - (\phi_{i+1,j+1} + \phi_{i+1,j})\right)^2 - \frac{\delta t}{\delta x}\cdot \frac{1 - i \varepsilon}{8}\,(\phi_{i+1,j+1} - \phi_{i+1,j})^2 \\ \nonumber \\
S[\bigtriangleup_{i,j}] &= \frac{\delta x}{\delta t}\, \cdot \frac{1 + i \varepsilon}{8}\,\left(2 \phi_{i+1,j+1} - (\phi_{i,j+1} + \phi_{i,j})\right)^2 - \frac{\delta t}{\delta x}\cdot \frac{1 - i \varepsilon}{8}\,(\phi_{i,j+1} - \phi_{i,j})^2\,.
\end{align}
Then the total lattice action can be written as
\begin{equation}
\label{E:triangleaction1}
S_{\text{lattice}}[\{\phi_{i,j}\}]= \sum_{i,j} \big( S[\bigtriangledown_{i,j}]  + S[\bigtriangleup_{i,j}]  \big)\,.
\end{equation}

Next, we generalize our construction to account for a left-moving, accelerating mirror.  We take the mirror to have unit acceleration to the left so that its worldline is
\begin{equation}
(t(\tau), x(\tau)) = (\sinh(\tau),  1-\cosh(\tau))\,.
\end{equation}
If we were to plot this line on top of Figure~\ref{Fig:newtriangulation1}, it would intersect with our triangulation.  So we need to modify the triangulation to accommodate the new boundary.

\newpage 
\begin{figure}[h!]
\begin{center}
\includegraphics[scale=.55]{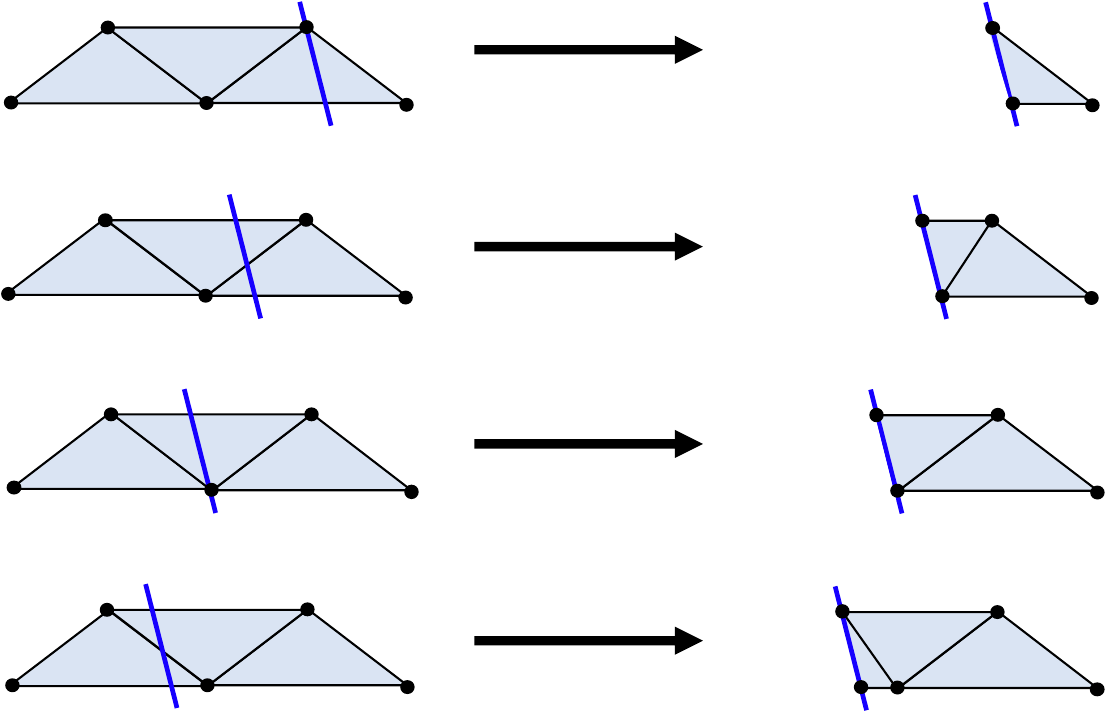}
\end{center}
\caption{A local depiction of the replacement rules for accommodating a mirror (the dark blue line) accelerating to the left.  The mirror's worldline is depicted as a straight line since for small enough triangulations the worldline appears as linear relative to the scale of individual triangular plaquettes. \label{fig:localreplacement}}
\end{figure}

$$$$
$$$$

\begin{figure}[h!]
\begin{center}
\includegraphics[scale=.38]{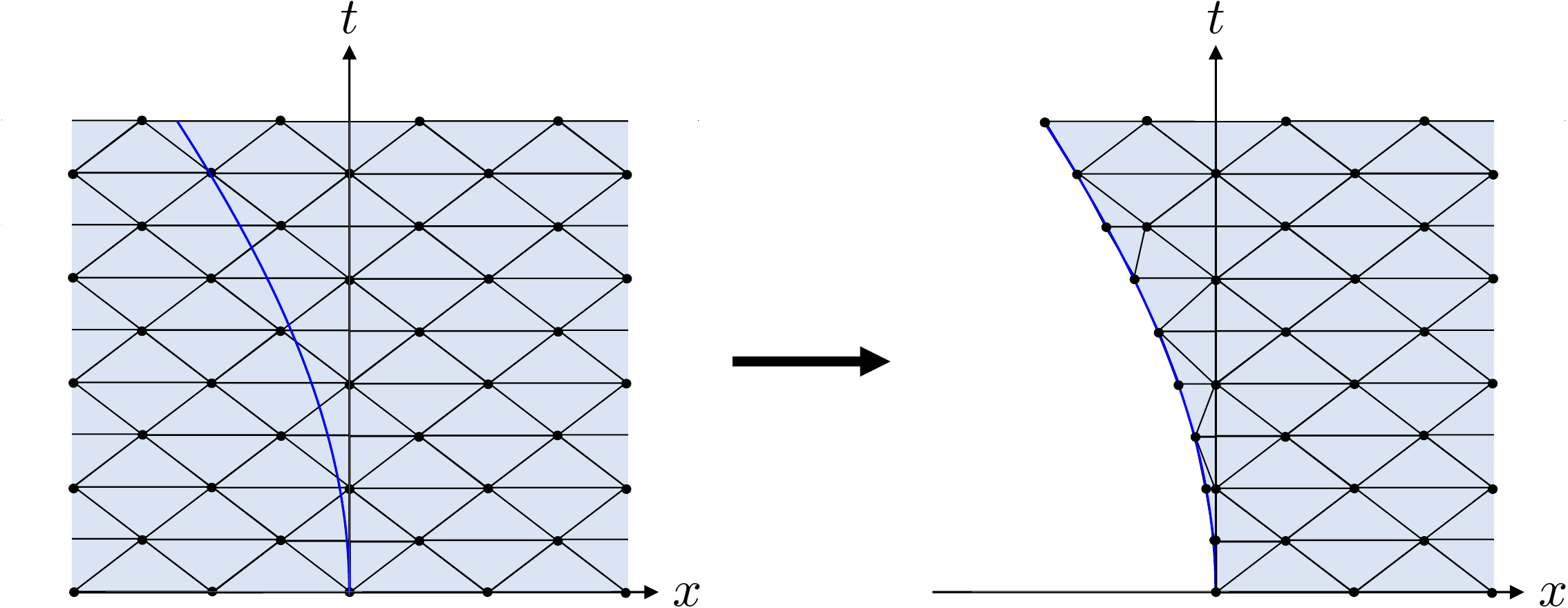}
\end{center}
\caption{The replacement rules applied to the entire triangular lattice. \label{fig:fullreplacement}}
\end{figure}

\newpage

At a formal level, let $\mathcal{T}$ be the space of triangulations of subsets of $\mathbb{R}^2$ for which each triangle has one edge parallel to the $x$-axis.  Our triangulation $\{\bigtriangledown_{i,j}\,, \bigtriangleup_{k,\ell}\}$ belongs to $\mathcal{T}$, and we would like a `replacement rule' which is a map $\mathcal{R} : \mathcal{T} \to \mathcal{T}$ such that $\mathcal{R}(\{\bigtriangledown_{i,j}\,, \bigtriangleup_{k,\ell}\}) = \{\bigtriangledown_{a,b}'\,, \bigtriangleup_{c,d}'\}$ is a new triangulation appropriate for the moving mirror boundary.  The replacement rule tells us which triangles we should throw away, and how we should modify triangles which intersect the mirror worldline.  Since our boundary is accelerating to the left, it suffices to employ the local replacement rules in Figure~\ref{fig:localreplacement}.  Here the replacement rules are shown locally when the worldline of the mirror (the dark blue line) intersects a triangular plaquette; we are to apply this replacement rule everywhere.  Indeed, the local replacement rules induces the map $\mathcal{R}(\{\bigtriangledown_{i,j}\,, \bigtriangleup_{k,\ell}\}) = \{\bigtriangledown_{a,b}'\,, \bigtriangleup_{c,d}'\}$ on the entire triangulation depicted in Figure~\ref{fig:fullreplacement}.

Observe that most of the triangles in the $x \geq 0$ region are unaltered.  For the altered triangles, we need to replace their contribution to the action in~\eqref{E:triangleaction1}.  Many of the triangles are removed entirely in our replacement procedure (i.e.~if they are to the left of the boundary) and so do not contribute to the new lattice action at all.  The other altered triangles are deformed; for instance, consider triangles of the general form
\begin{align}
\label{E:DownTriangle}
    \includegraphics[scale=.4, valign = c]{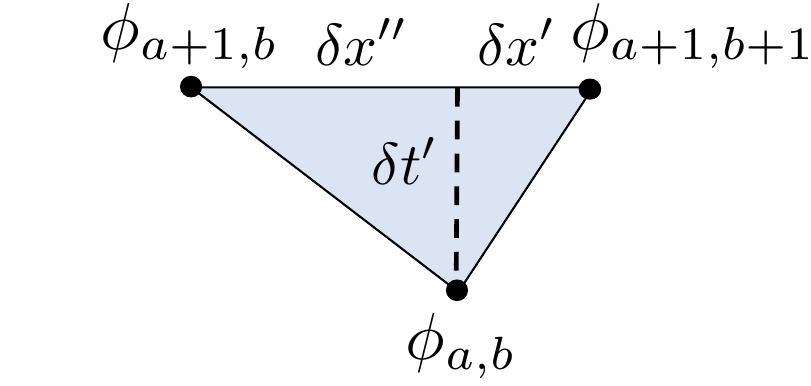}
\end{align}
and also
\begin{align}
\label{E:UpTriangle}
    \includegraphics[scale=.4, valign = c]{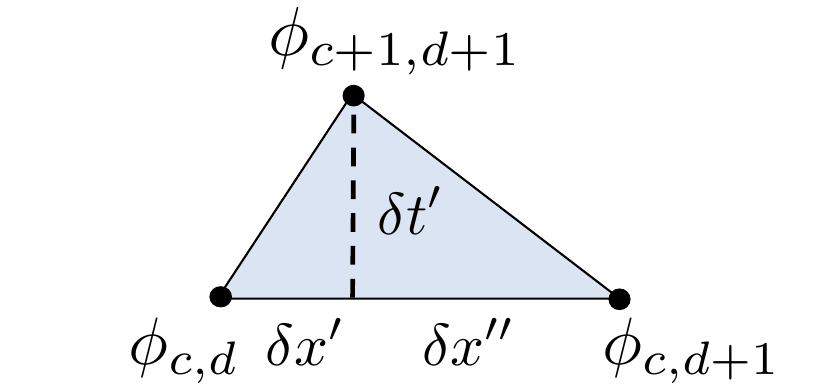}
\end{align}
where we have labeled their vertices with field degrees of freedom.  Then the lattice action for each of these triangles is, respectively,
\begin{align}
S[\bigtriangledown_{a,b}'] &= \frac{1}{\delta t' (\delta x' + \delta x'')}\, \frac{1 + i \varepsilon}{4}\,\Big( \delta x'' \, \phi_{a+1,b+1} + \delta x' \, \phi_{a+1,b} - (\delta x' + \delta x'') \, \phi_{a,b}\Big)^2 \nonumber \\
&\qquad \qquad \qquad \qquad \qquad \qquad \qquad \qquad \quad \,\,\, - \frac{\delta t'}{\delta x' + \delta x''}\cdot \frac{1 - i \varepsilon}{4}\,(\phi_{a+1,b+1} - \phi_{a+1,b})^2\,, \\ 
S[\bigtriangleup_{c,d}'] &= \frac{1}{\delta t' (\delta x' + \delta x'')}\, \frac{1 + i \varepsilon}{4}\,\Big( \delta x'' \, \phi_{c,d} + \delta x' \, \phi_{c,d+1} - (\delta x' + \delta x'') \, \phi_{c+1,d+1}\Big)^2 \nonumber \\
&\qquad \qquad \qquad \qquad \qquad \qquad \qquad \qquad \qquad \quad - \frac{\delta t'}{\delta x' + \delta x''}\cdot \frac{1 - i \varepsilon}{4}\,(\phi_{c,d+1} - \phi_{c,d})^2\,.
\end{align}
Our original triangles are a special case of the above with $\delta x' = \delta x'' = \delta x$.  By summing over the actions of all of the triangles in our modified triangulation $\{\bigtriangledown_{a,b}'\,, \bigtriangleup_{c,d}'\}$, we obtain our desired lattice action.  To impose Dirichlet boundary conditions on the mirror, we simply set the field elements along the mirror boundary equal to zero.  In total, we can express the lattice action as
\begin{equation}
S_{\text{lattice}}^{\text{mirror}}[\{\phi_{i,j}\}] = \left(\sum_{\bigtriangledown_{a,b}'} S[\bigtriangledown_{a,b}'] + \sum_{\bigtriangleup_{c,d}'} S[\bigtriangleup_{c,d}']\right)\Bigg|_{\text{ boundary }\phi\text{'s} \,=\, 0}\,.
\end{equation}

Examining the triangulation on the right-hand side of Figure~\ref{fig:fullreplacement}, we see that the corresponding lattice action propagators from the past to the future will instantiate Hilbert space isometries.  Moreover the future to past propagators are projections.

The fact that the theory defined with a lattice cutoff is isometric but not unitary is in general accord with the continuum observation of section 2.1 that any finite resolution detector will not see unitary evolution. In principle it would be interesting to understand the continuum limit of this FEM latticization in enough detail to reproduce the numerical value~\eqref{snt} of the entanglement entropy produced by an accelerating mirror.

\subsection{The physical subspace}
\label{subsec:physical}

Here we discuss some physical consequences of having time evolution be instantiated by an isometry which is not unitary.
% So far, we have mostly considered the setting of quantum field theory coupled to an expanding background.  We have demonstrated how mandating a spatial UV cutoff $\ell$ on each Cauchy slice, where $\ell$ has fixed proper length, forces us to consider isometries between spatial slices.  In a cosmological setting, quantum field theory on a classical expanding background should be regarded as an effective field theory, in the regime where the matter stress tensor is not strong enough to backreact on the geometry.  Nonetheless, our basic approach makes sense in this setting as well.  In the discussion below, we will keep our analysis sufficiently general that it may pertain to any and all of the above settings.

Suppose we have a sequence of Hilbert spaces $\mathcal{H}_{t_0}, \mathcal{H}_{t_1}, \mathcal{H}_{t_2}, ...$ with increasing dimensions $d_0 < d_1 < d_2 < \cdots$.  Moreover, let our time evolution be instantiated by an isometric but non-unitary propagator $\mathcal{K}_{j \leftarrow i}$ which maps $\mathcal{H}_{t_i} \to \mathcal{H}_{t_j}$ for $i < j$.  The initial Hilbert space $\mathcal{H}_{t_0}$ has the smallest dimension, $d_0$.  We identify $\mathcal{H}_{t_0}$ with the `physical' Hilbert space $\mathcal{H}_{\text{phys}}$, namely $\mathcal{H}_{t_0} \simeq \mathcal{H}_{\text{phys}}$, for reasons that we explain shortly.

For concreteness, imagine that have an initial state $|\Psi_0\rangle$ in $\mathcal{H}_\text{phys}$, and evolve it according to the propagator.  If we evolve it via $\mathcal{K}_{j \leftarrow 0}$, then $|\Psi_j\rangle := \mathcal{K}_{j \leftarrow 0} |\Psi_0\rangle$ is in $\mathcal{H}_j$.  Suppose we act on $|\Psi_j\rangle$ by an operator $O_j$ which takes $\mathcal{H}_j$ to itself; that is, we obtain $|\Psi_j'\rangle := O_j |\Psi_j\rangle$ which is still in $\mathcal{H}_j$.  We can ask: is there an alternative initial state $|\Psi_0'\rangle$ in $\mathcal{H}_{\text{phys}}$ such that $|\Psi_j'\rangle = \mathcal{K}_{j \leftarrow 0}|\Psi_0'\rangle$?  At a physical level, we are asking if there are any initial conditions which could have evolved into $|\Psi_j'\rangle$.

Let $\mathcal{H}_{\text{phys}, \,t_j}$ denote the image of $\mathcal{H}_{\text{phys}} \simeq \mathcal{H}_{t_0}$ under evolution by the propagator $\mathcal{K}_{j \leftarrow 0}$.  Then the answer to the above question is that there exists such a $|\Psi_0'\rangle$ if and only if $|\Psi_j'\rangle$ is in $\mathcal{H}_{\text{phys},\, t_j}$.  Since $\mathcal{H}_{\text{phys},\, t_j}$ is a proper subspace of $\mathcal{H}_{t_j}$ with dimension $d_0$, we see that most states $|\Psi_j'\rangle$ do not have antecedent states $|\Psi_0'\rangle$.

The above is essentially stipulating that the sequence of bijections
\begin{equation}
\mathcal{H}_{\text{phys}} \xrightarrow{\mathcal{K}_{1 \leftarrow 0}} \mathcal{H}_{\text{phys},\,t_1} \xrightarrow{\mathcal{K}_{2 \leftarrow 1}}  \mathcal{H}_{\text{phys},\,t_2} \xrightarrow{\mathcal{K}_{3 \leftarrow 2}} \cdots 
\end{equation}
comprises the physical evolution of physical states.  In fact, from this point of view, the evolution restricted to the physical subspaces is completely unitary.  Even though $\mathcal{H}_{t_j}$ is larger in dimension than $\mathcal{H}_{\text{phys}} \simeq \mathcal{H}_{t_0}$, this is in effect illusory since the only physically accessible states in $\mathcal{H}_{t_j}$ are those in $\mathcal{H}_{\text{phys},\, t_j}$.  A consequence is that the physical algebra of observables on $\mathcal{H}_{t_j}$ is composed of precisely those observables that map $\mathcal{H}_{\text{phys},\, t_j}$ to itself (or stronger, those observables which are the identity on the orthogonal complement of $\mathcal{H}_{\text{phys},\, t_j}$ in $\mathcal{H}_{t_j}$).

\subsection{Summary and comments}

Our perspective of restricting to physical subspaces recovers unitarity, but allows for the number of `apparent' degrees of freedom to increase in time.  There are parallels between the above and the `code subspace' in the error correction interpretation of AdS/CFT bulk reconstruction.  However, we note that we have no argument that our isometries above need to be ones which comprise a quantum error correction code robust to spatially local errors. It would be interesting to consider this point in the cosmological context. 

We conclude with comments about quantum fields on an expanding background.  Even for a scalar coupled to gravity in the continuum limit, not all states of the quantum field in the far future will correspond to antecedent states in the far past.  For instance, many configurations when evolved backwards will lead to singularities at finite time, past which we cannot evolve back further. One might conclude some of these late-time states could be ruled out once a suitable definition of the physical subspace is understood in this setting. But this could be too hasty: an alternate possibility is that the scalar effective field theory breaks down and more degrees of freedom are needed. Indeed in an analogous AdS setting this line of reasoning leads to the inclusion of black holes \cite{Pastawski:2015qua} rather than constraints on the boundary Hilbert space. This point is also worthy of further consideration.

\section{Embedding dS in AdS/CFT}
\label{Sec:embedding}

Quantum gravity in de Sitter space can be holographically realized by embedding it as an RS-type braneworld near the boundary of AdS \cite{Hawking:2000da}.
On the other hand, the holographic structure of AdS itself is captured by a quantum error-correcting code  of  a tensor network \cite{Almheiri:2014lwa,Pastawski:2015qua}. In this section we consider in particular the HaPPY code \cite{Pastawski:2015qua} for AdS$_3$ and ask what structure it imparts to  the boundary dS$_2$ braneworld. We indeed find, in the simplest adaptation, that the number of dS braneworld bits grows with  the spacetime expansion. In this model we find that time evolution on dS$_2$ is a non-unitary isometry which forms a quantum error-correcting code.  Moreover this evolution reproduces the logarithmic growth of the entanglement entropy $S_{ent}$ found in the continuum analysis in~\eqref{dse}. 

We first briefly review the relevant features of the dS$_2 \in \,$AdS$_3$  braneworld construction in \cite{Hawking:2000da}. The metric  for an AdS$_3$ geometry with radius $L$ can be written
\be \label{dsv}ds_3^2=L^2\left(dr^2-\sinh^2 r {dt^+dt^-\over \cos^2 t}\right)\,,\ee
where $2t=t^++t^-$ and  $-{\pi \over 2}<t<{\pi \over 2}$.  The geometry can be terminated at a 1-brane at large  fixed $r_B \gg 1$.
The Israel matching condition requires the tension $T$ to obey
\be  T ={\coth r_B\over 4{\pi} G_3L} ,\ee
where $G_3$ is the 3D Newton constant. 
The induced metric on the dS$_2$  worldbrane is then 
\be \label{rta}ds_2^2=-R^2(t)dt^+dt^-,~~~~R(t)={\ell \over \cos t}\,, \ee
with  dS$_2$ radius 
\be \ell=L\sinh r_B\,.\ee
This terminated AdS$_3$ geometry is dual \cite{Hawking:2000da} to a CFT$_2$ with Brown-Henneaux central charge 
\be\label{kkl} c={3L \over 2G_3}  \ee
in  the dS$_2$ spacetime coupled to 2D gravity at a cutoff $L$.\footnote{Reference \cite{Hawking:2000da} considers a double-cover geometry in which the dS$_2$-brane bounds two AdS$_3$ regions, carries twice the central charge and has twice the action.  Here we take a quotient so that there is only one AdS$_3$ region and the central charge is not multiplied by two.}
%half the central charge.
 The 2D gravitational action is induced by the cutoff CFT$_2$ and has an effective Newton constant 
\be {1 \over G_2}= {L \log \ell \over G_3}\,.\ee
Since the dS$_2$ horizon consists of 2 points,  the  area law for dS$_2$ entropy gives
\be S_{dS}={Area \over 4 G_2}= {L \log \ell \over 2G_3}\,.\ee
On the other hand, using the fact that $t, \phi$ are dS$_2$ vacuum coordinates, inserting the central charge \eqref{kkl} into formula \eqref{ddd} for the   entanglement entropy gives 
\be S_{ent}={c \log \ell \over 3}=S_{dS}\,.\ee
Hence in this setup the macroscopic area law is explained by microscopic entanglement.  This is apparently  a general feature of any theory in which gravity is induced \cite{Jacobson:1994iw,Fiola:1994ir,Donnelly:2012st}, for both cosmological and black hole horizons.

\begin{figure}
\begin{center}
\includegraphics[scale=.55]{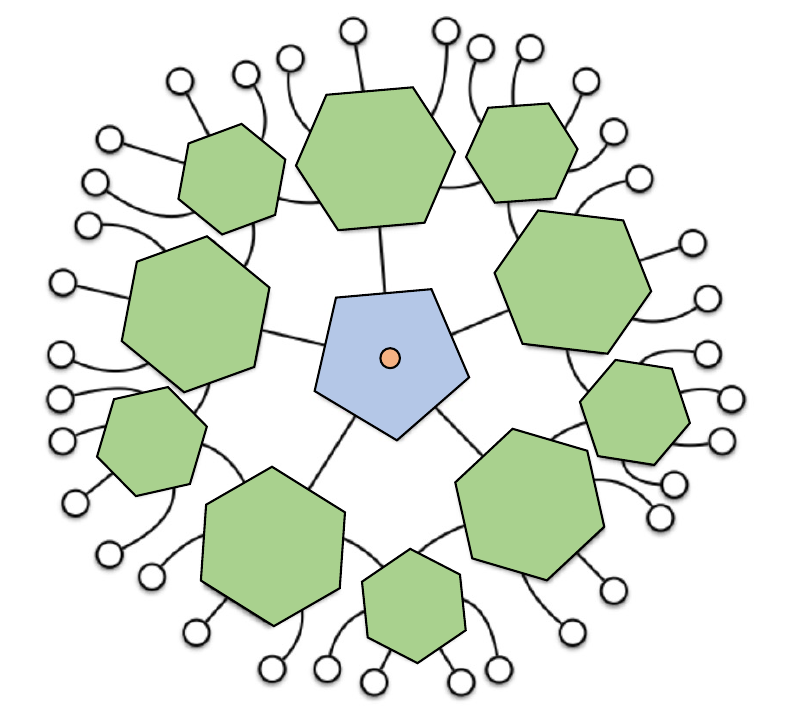}
\end{center}
\caption{The HaPPY one qubit hexagon code~\cite{Pastawski:2015qua}.  One logical qubit (the central, orange-colored node)  is the sixth leg of a pentagonally-shaped perfect tensor, with each of the outputs of the pentagon feeding into hexagonally-shaped perfect tensors.  The hexagonally-shaped perfect tensors feed into other hexagonally shaped perfect tensors.  The white nodes at the periphery correspond to the encoded state.   \label{Fig:Happy1}}
\end{figure}

Now we briefly review the relevant features of AdS$_3$ tensor network holography as presented in \cite{Almheiri:2014lwa,Pastawski:2015qua}.  For our illustrative purposes, it suffices to consider the so-called one qubit hexagon code\cite{Pastawski:2015qua}, in which a single bulk qubit at the origin is holographically mapped to an $n$-qubit state at the boundary; see Figure~\ref{Fig:Happy1}.  Global coordinates for AdS$_3$ are 
%\be ds^2=-(R^2+L^2)d\tau^2+{dR^2 \over R^2+L^2} +R^2d\phi^2.\ee
\be ds^2=-(1 + R^2/L^2)d\tau^2+{dR^2 \over 1 + R^2/L^2} +R^2d\phi^2.\ee
A spatial  slice of fixed global time $\tau$  has the geometry of the hyperbolic plane. This has a hexagonal  tiling   in which 4 hexagons meet at every corner of every hexagon.  Qubits are assigned to their edges. In the one qubit hexagon code, a pentagon is inserted at the origin.  The `logical' qubit placed in the center  of this pentagon models degrees of freedom of a bulk quantum field. At some large radius $R_B\gg L$ there is  a boundary (drawn to miss vertices) which  gives  of order  \be \label{nr} n\sim {R_B \over L} \ee boundary qubits.   One  seeks a  holographic duality map from a bulk state of the logical qubit near the center of AdS$_3$  to an $n$-qubit boundary state. This is given by an $n+1$ index tensor $T_{i_1...i_n; j}$\,,
% \be \label{hpp} T_{i_1...i_n; j} =\langle i_1...i_n;j\rangle \ee
where all indices take the value $(0,1)$ indicating the component of a qubit. $T_{i_1...i_n; j}$  is built from a tensor network of 
6-index tensors $T_{i_1,...i_6} $ associated to each polygon.  For hexagons the indices $i_k$ all corresponds to the states of qubits on the edges, while for the pentagon the 6th index denoted $j=i_6$ is the logical  qubit at the center. Contracting the product of all these tensors over the edge indices then gives the map $T_{i_1...i_n; j}$ from bulk to boundary states (i.e.~from the $j$ index to the $i_1,...,i_n$ indices). That is, the remaining free indices define a map from the  bulk logical qubit to $n$ boundary qubits.   In \cite{Pastawski:2015qua}, an explicit choice of $T_{i_1...i_6} $ was presented with a number of beautiful properties including that $T_{i_1...i_n; j}$ is not only  an isometry but  a quantum error-correcting code. This gives the sought-after holographic map from the single-qubit bulk Hilbert space to the $n$-qubit boundary Hilbert space.  The construction easily generalizes to a much larger $k$-qubit bulk Hilbert space corresponding to the low-energy sector of the boundary CFT\cite{Pastawski:2015qua}.

The boundary Hilbert space may be enlarged outward  by increasing $R_B$ which adds more hexagons.  This defines an isometric embedding $T(R)$ of the $1$-qubit logical code subspace into ever-larger Hilbert spaces, as well as isometric maps between ever larger boundary Hilbert spaces. $T(R)$ corresponds to inverse RG flow which embeds states in a low-energy effective Hilbert space into the ever-larger Hilbert space which appears as the cutoff is removed.\footnote{For a useful recent discussion of the well-known  connection between RG flow and  quantum error-correcting codes see \cite{Furuya:2021lgx}.}

Now let us combine the observation of \cite{Hawking:2000da,Almheiri:2014lwa,Pastawski:2015qua} to  make a tensor network type model for dS$_2$.  
 Instead of drawing an $S^1$ boundary at large radius $R_B$ in a spatial section of AdS$_3$, we draw a dS$_2$  boundary in  AdS$_3$ at radius $R(t)$ as given in \eqref{rta},
 throwing away all  qubits outside this region.
In the expanding region with dS$_2$ time $t>0$, the number of boundary qubits will grow with the radius. The time evolution of the  boundary state therefore cannot be unitary. Instead it is given by the isometry
\be |\Psi(t)\rangle =  T(R(t))\,|\Psi(0)\rangle\,.\ee
We see that the holographic and isometric relation between boundary state at different radii $R$ in static AdS$_3$ directly translates into a holographic and isometric but non-unitary relation between bulk states in dS$_2$ at different times $t$. 
In  this toy model the quantum  state of the expanding universe evolves according to a quantum error-correcting code. 

This picture is consistent with the continuum computation of dS subregion entanglement entropy $S_{ent}$ given in \eqref{rgg}. Entanglement entropy of a subregion of the boundary circle in this tensor network model goes as the log of the number of subregion qubits \cite{Pastawski:2015qua}.  One therefore finds 
\be S_{ent} \sim \log R(t)  \ee 
in qualitative agreement with \eqref{rgg}. In an as-yet-unknown more advanced tensor network model which has a continuum limit reproducing the AdS$_3$ central charge, one would expect to reproduce both the prefactor in \eqref{rgg} and the microscopic derivation of dS$_2$ entropy from quantum entanglement. 

\section{dS/CFT}
\label{Sec:dSCFT}

The proposal of  this paper that time evolution in an expanding universe is isometric but not unitary both implies and is a refinement of the dS/CFT correspondence \cite{Strominger:2001pn}, particularly in the formulation of \cite{Strominger:2001gp} with time evolution dual to inverse RG flow. 

Let us first recall the basic idea of dS/CFT. 
The holographic dual of an expanding patch of dS$_{d+1}$  is conjectured to be a CFT$_d$ living on the boundary of the spacetime at future null infinity (${\cal I}^+$).\footnote{We consider here only one asymptotic boundary as apparently is the case in our own universe. However in perfect dS there is both a past and future boundary, perhaps indicating a type of thermofield double.} This basically turns the AdS/CFT correspondence sideways, with time replacing the radius as the holographically emergent dimension. The bulk-to-boundary dictionary, as further developed in \cite{Maldacena:2002vr,Harlow:2011ke,Ng:2012xp}, is a (subtle) $r\to t$ rotation of the AdS/CFT dictionary. Realistic cosmologies are dual to boundary field theories which are conformally invariant only in the extreme UV. This corresponds to a dS phase in the infinite future, in accordance with expectations from astronomical observations \cite{SupernovaCosmologyProject:1998vns}. Non-trivial RG flows can describe non-dS cosmologies as in the current epoch of our universe. The beginning of the universe corresponds to the extreme IR. A boundary theory which flows to a trivial fixed point is dual to a big bang cosmology with nothing at all before the beginning of time. On the other hand, a nontrivial CFT in the extreme IR corresponds to a primordial inflationary epoch of infinite duration.

Obviously time evolution cannot be unitary in dS/CFT.  It is dual to inverse RG flow and therefore maps smaller effective Hilbert spaces to larger ones. New degrees of freedom are continuously added to the universe as it expands, rather than all at once during the big bang. Nothing gradually evolves to something. 

The discussion in~\cite{Strominger:2001gp,Strominger:2001pn} focused on computations of correlation functions and did not address the evolution of quantum states in an expanding universe. Here we see that isometries are just what is needed to describe this expansion.  Moreover it is understood how inverse RG flow gives rise to isometries in the language of tensor networks (see e.g.~\cite{white1992density, white1993density, verstraete2004density, daley2004time, schollwock2005density, shi2006classical, levin2007tensor, Vidal:2007hda, gu2008tensor, xie2012coarse, haegeman2013entanglement, evenbly2014class, cotler2019renormalization, cotler2019entanglement}) and quantum error correction~\cite{Kim:2016wby, Furuya:2020tzv, Furuya:2021lgx}. Hence we propose here as a refinement of dS/CFT that the time evolution of states is given by isometries, and that these isometries are generated by inverse RG flow of the dual field theory at ${\cal I}^+$.

\section{Discussion}
\label{Sec:Discuss}

We conclude with some additional remarks.

%miscellaneous remarks.

As we stated earlier, a conceptual argument in favor of time evolution being instantiated by isometries in expanding universes is that the volume  of space grows as a function of time but the Planck length stays fixed.  There is a more holographic analog of this for universes in which the horizon of each observer is expanding, such as our own universe in its present phase.  Suppose that an observer were to identify the area of his cosmological horizon with the number of degrees of freedom which describe the universe; if this is increasing for all observers, then the number of degrees of freedom is increasing with time.  Thus even a holographic perspective appears to necessitate some form of isometric but non-unitary time evolution.

The exception is perfect de Sitter space, where  the area of an observer's horizon does not change and there are two asymptotic boundaries.   This appears to run counter to the necessity of isometric but non-unitary time evolution.  Nonetheless, the rules of de Sitter holography are presently far from clear, and in any case we do not currently live in a static patch.

In subsection~\ref{subsec:physical} we discussed the role of the `physical subspace' in a universe with isometric time evolution.  We saw that the physical subspace is mapped unitarily to another physical subspace of the same dimension.  In this sense, isometric time evolution is just unitary evolution when we restrict to the physical subspaces.  So what is the role of the total ambient Hilbert space, which increases in size as time advances?  An analogy with gauge theory may be appropriate.  A gauge theory is not strictly local due to gauge constraints (i.e.~it does not have a tensor product structure), but we can render it as local by embedding it in a larger ambient Hilbert space.  However, the physical states only lie in the gauge-invariant (physical) subspace.  Turning back to theories in expanding universes, we need a progressively larger ambient Hilbert space as time advances to keep rendering the theory as local.  Said a different way, while we could instead choose to describe an expanding universe solely in terms of unitary evolution of the physical subspace, the dynamics may be non-local; to make the dynamics look local we may require a larger ambient Hilbert space whose size increases with time, and as such need isometric evolution.  This suggests a possible tension between unitarity and locality in expanding universes: we can either describe evolution as unitary and non-local, or isometric and local.

An obvious question in light of our hypothesis about expanding geometries is: how should we think about contracting geometries?\footnote{See \cite{Barbon:2011ta} for an interesting analysis of crunch geometries in AdS holography.} We do not have much to say here. This is a rather different question because excitations are blueshifted in contracting geometries such as a big crunch or the interior of a black hole. One is thereby driven out of the region of validity of effective field theory and eventually into the realm of strongly coupled quantum gravity.  If we ignore this and na\"{i}vely time reverse the propagator in 
\eqref{E:kernel1} the isometry becomes a projection. Perhaps the forward arrow of time is correlated with the isometric direction so that we always perceive geometries as expanding. This could leads to a multi-history picture as in \cite{gell1994time} or a final state projection as in \cite{Horowitz:2003he}; see also~\cite{Cotler:2018skm}. We leave this to future study. 

Throughout the present paper, our analyses pointed to interesting structures in the encodings provided by isometric time evolution in quantum field theory and hypothetically in quantum gravity.  Understanding the precise information-theoretic properties of these encodings appears to be an interesting avenue for future pursuit.

\subsection*{Acknowledgements}
We thank Netta Engelhardt, Hao Geng, Felipe Hern\'{a}ndez, Laura Niermann, Tobias Osborne, Djordje Radicevic, Daniel Ranard, Tadashi Takayanagi and Frank Wilczek for valuable discussions.
This work is supported by  the Harvard Society of Fellows, the Gordon and Betty Moore and Templeton Foundations via the Black Hole Initiative and  the Department of Energy under grant DE-SC0007870.

\appendix

\section{$\mathcal{K}\mathcal{K}^\dagger$ is a projector for the free scalar on a triangle}
\label{App:isom1}

In section~\ref{Subsec:pathintegral1} we established that the kernel $\mathcal{K} : \mathcal{H}_0 \to \mathcal{H}_1 \otimes \mathcal{H}_2$, defined via its position-space representation in~\eqref{E:kernel1} (see also~\eqref{E:singletriangle2} and~\eqref{E:const1}), satisfies $\mathcal{K}^\dagger \mathcal{K} = \mathds{1}_{\mathcal{H}_0}$.  Let us now consider $\mathcal{K}\mathcal{K}^\dagger$ which evolves the system from $t = 1$ to $t = 0$, and then back to $t = 1$.  This satisfies $(\mathcal{K}\mathcal{K}^\dagger) (\mathcal{K}\mathcal{K}^\dagger) = \mathcal{K}\mathcal{K}^\dagger$, and so it is a projector; thus $\mathcal{K}$ is an isometry.  We would like to establish that this isometry is not a unitary, namely that $\mathcal{K}\mathcal{K}^\dagger \not = \mathds{1}_{\mathcal{H}_1 \otimes \mathcal{H}_2}$.  For this it suffices to show that $\mathcal{K}\mathcal{K}^\dagger$ has a nontrivial nullspace.  

Returning to the position-space representation, $\mathcal{K}\mathcal{K}^\dagger$ corresponds to
\begin{align}
\label{E:KKdag1}
&\int d\phi_0 \, K(\phi_1, \phi_2\,;\,\phi_0) \, K^*(\phi_1', \phi_2'\,;\,\phi_0) \nonumber \\
& \quad = \frac{1}{4\pi} \, \exp\left(- \frac{1}{16\,\varepsilon}\frac{\delta x}{\delta t}\left((\phi_1 + \phi_2) - (\phi_1' + \phi_2')\right)^2 - \frac{i}{8} \frac{\delta t}{\delta x}\left((\phi_1 - \phi_2)^2 - (\phi_1' - \phi_2')^2\right) - \varepsilon\, f(\phi_1, \phi_2, \phi_1', \phi_2')\right) \nonumber \\
& \quad  \sim \sqrt{\frac{\delta t}{\delta x}\frac{\varepsilon}{\pi}}\, \delta\!\left((\phi_1 + \phi_2) - (\phi_1' + \phi_2')\right)\, e^{\frac{i}{8}  \frac{\delta t}{\delta x}\left((\phi_1 - \phi_2)^2 - (\phi_1' - \phi_2')^2\right) - \varepsilon\, f(\phi_1, \phi_2, \phi_1', \phi_2')}
\end{align}
where
\begin{align}
f(\phi_1, \phi_2, \phi_1', \phi_2') =  \frac{1}{16} \left\{\frac{\delta x}{\delta t}\left((\phi_1 + \phi_2) - (\phi_1' + \phi_2')\right)^2 + 2 \,\frac{\delta t}{\delta x} \left((\phi_1 - \phi_2)^2 + (\phi_1' - \phi_2')^2\right)\right\}\,.
\end{align}
The `$\sim$' in~\eqref{E:KKdag1} designates that we are taking the limit of small $\varepsilon$.  To interpret~\eqref{E:KKdag1} let us integrate our kernel against a normalizable wavefunction $\Psi(\phi_1, \phi_2)$, that is:
\begin{equation}
\sqrt{\frac{\delta t}{\delta x}\frac{\varepsilon}{\pi}} \int d\phi_1 \, d\phi_2 \, \delta\!\left((\phi_1 + \phi_2) - (\phi_1' + \phi_2')\right)\, e^{-\frac{i}{8} \frac{\delta t}{\delta x}\left((\phi_1 - \phi_2)^2 - (\phi_1' - \phi_2')^2\right) - \varepsilon \,f(\phi_1, \phi_2, \phi_1', \phi_2')  }\,\Psi(\phi_1, \phi_2)\,.
\end{equation}
Here it will be useful to consider the change of variables $\chi_1 = \phi_1 + \phi_2$, $\chi_2 = \phi_1 - \phi_2$, $\chi_1' = \phi_1' + \phi_2'$, $\chi_2' = \phi_1' - \phi_2'$.  The measure picks up a factor of $\frac{1}{2}$ upon changing $\phi_1, \phi_2$ to $\chi_1, \chi_2$.  Since we are also changing $\phi_1', \phi_2'$ to $\chi_1', \chi_2'$ and ultimately want our kernel-integrated wavefunction to be $L^2$-normalized with respect to the $d\chi_1' \, d\chi_2'$ measure, we multiply by an additional $\frac{1}{\sqrt{2}}$.  Then our integral becomes
\begin{equation}
\label{E:chivariables1}
\sqrt{\frac{\delta t}{\delta x}\frac{\varepsilon}{8\pi}} \int d\chi_1 \, d\chi_2 \, \delta(\chi_1 - \chi_1' )\, e^{-\frac{i}{8} \frac{\delta t}{\delta x}\left(\chi_2^2 - \chi_2'^2\right) - \varepsilon \, f(\chi_1, \chi_2, \chi_1', \chi_2') }\,\Psi(\chi_1, \chi_2)\,.
\end{equation} 
Performing the $\chi_1$ integral, we obtain
\begin{equation}
\label{E:chivariables2}
\sqrt{\frac{\delta t}{\delta x}\frac{\varepsilon}{8\pi}} \,\,e^{-\frac{\varepsilon - i}{8} \frac{\delta t}{\delta x}\,\chi_2'^2}\int d\chi_2 \, e^{-\frac{\varepsilon +  i}{8} \frac{\delta t}{\delta x}\,\chi_2^2}\,\Psi(\chi_1', \chi_2)
\end{equation}
Interestingly, there is an entire subspace of $\Psi$'s for which the right-hand side vanishes in the $\varepsilon \to 0$ limit.  For instance, 
\begin{equation}
\Psi(\chi_1, \chi_2) = \frac{1}{\sqrt{2\pi}} \, \exp\left(- \frac{1}{4}\, \chi_1^2 - \frac{1}{4} \, \chi_2^2\right)
\end{equation}
has the property that
\begin{equation}
\mathcal{K}\mathcal{K}^\dagger |\Psi\rangle = 0
\end{equation}
This itself establishes that $\mathcal{K}\mathcal{K}^\dagger$ has a non-trivial nullspace,  and accordingly $\mathcal{K}$ is an isometry which is not a unitary.

It is instructive to consider a wider class of examples.
%\textcolor{red}{[Can we construct a more detailed analysis of the subspace?]}
Suppose that $\Psi(\chi_1, \chi_2)$, which we have taken to be normalizable, has the form
\begin{equation}
\Psi(\chi_1, \chi_2) = \exp\!\left(\sum_{m,n = 0}^\infty c_{m,n} \, \chi_1^m \chi_2^n\right)
\end{equation}
for the $c_{m,n}$'s complex.  Then $\mathcal{K}\mathcal{K}^\dagger |\Psi\rangle = 0$ in the $\varepsilon \to 0$ limit if $c_{0,2} \not = \frac{i}{8} \frac{\delta t}{\delta x}$.  By contrast, if $c_{0,2} = \frac{i}{8} \frac{\delta t}{\delta x}$, then $\mathcal{K}\mathcal{K}^\dagger |\Psi\rangle \not = 0$, in particular
\begin{equation}
\langle \chi_1, \chi_2 | \mathcal{K}\mathcal{K}^\dagger |\Psi\rangle = \exp\!\left(\sum_{m = 0}^\infty  c_{m,0} \, \chi_1^m + \frac{i}{8} \frac{\delta t}{\delta x} \,\chi_2'^2 \right)\,.
\end{equation} 
So such $|\Psi\rangle$'s are not in the nullspace of $\mathcal{K}$.  Note that the right-hand side of the above equation is not normalizable; we address this below in Appendix~\ref{App:spaces}.
% Thus we have shown that $\mathcal{K}\mathcal{K}^\dagger$ has a non-trivial nullspace, and accordingly $\mathcal{K}$ is an isometry which is not a unitary.

% \textcolor{red}{[Emphasize that $\mathcal{K}\mathcal{K}^\dagger$ has a non-trivial nullspace]}

\section{Function spaces accommodating the $i\varepsilon$ prescription}
\label{App:spaces}

In our analysis of isometries in section~\ref{Subsec:pathintegral1}, the $i \varepsilon$ prescription was essential to get sensible answers.  Moreover, the $i\varepsilon$ plays a more dramatic role in our analyses than in ordinary quantum field theory.  We begin by considering an illuminating example.  Consider again a free scalar field on a triangle, with the propagator given by~\eqref{E:kernel1} (which uses~\eqref{E:singletriangle2} and~\eqref{E:const1}).  Suppose we have a normalizable wavefunction $\Psi(\phi_0)$ on the vertex in the past, and evolve it by the propagator.  This yields
\begin{align}
\label{E:propPsi0}
&\int d\phi_0 \, K(\phi_1, \phi_2 \, ; \, \phi_0)\, \Psi(\phi_0) = \frac{1}{2}\left(\frac{\delta x}{\delta t}\frac{\varepsilon}{\pi^3}\right)^{1/4} e^{ - \frac{\delta t}{\delta x} \frac{\varepsilon + i}{8}\,(\phi_1-\phi_2)^2}\int d\phi_0 \, e^{- \frac{\delta x}{\delta t}\frac{\varepsilon - i}{8}\,\left(2 \phi_0 - (\phi_1 + \phi_2)\right)^2}\, \Psi(\phi_0)\,.
\end{align}
Note that the integral over $\phi_0$ on the right-hand side only depends on the combination $\phi_1 + \phi_2$ which we denote by $\chi_1$.  Letting $\chi_2 = \phi_1 - \phi_2$, we can rewrite the right-hand side of the above as
\begin{equation}
\label{E:wavefunctionexample1}
\widetilde{\Psi}_1(\chi_1) \cdot \frac{1}{\sqrt{2}}\left(\frac{\delta t}{\delta x} \, \frac{\varepsilon}{\pi}\right)^{1/4} e^{ - \frac{\delta t}{\delta x} \frac{\varepsilon + i}{8}\,\chi_2^2}\,
\end{equation}
where $\widetilde{\Psi}_1(\chi_1)$ is an $L^2$-normalized wavefunction.  Observe that the $\chi_2$ part of the above expression is itself an $L^2$-normalized wavefunction, albeit one with variance $\sim 1/\sqrt{\varepsilon}$.  A peculiar feature is that~\eqref{E:wavefunctionexample1} vanishes in the $\varepsilon \to 0$ limit, although if we take its $L^2$ norm for finite $\varepsilon$ and then take $\varepsilon \to 0$ we get $1$.  Apparently the order of limits matters here: we should compute the norm before taking the $\varepsilon \to 0$ limit.

As a very explicit example, let
\begin{equation}
\Psi(\phi_0) = \frac{1}{(2\pi)^{1/4}} \, e^{- \frac{1}{4}\,\phi_0^2}
\end{equation}
so that~\eqref{E:propPsi0} becomes
\begin{align}
\Psi(\phi_1, \phi_2) &= \frac{\left(\frac{\delta t}{\delta x}\, \frac{\varepsilon}{2\pi^2}\right)^{1/4}}{\sqrt{\frac{\delta t}{\delta x} + 2(\varepsilon - i)}} \,\, e^{- \frac{1}{8\left(\frac{\delta t}{\delta x} + 2 ( \varepsilon - i)\right)}\left[\left(\frac{\delta t^2}{\delta x^2}(\varepsilon + i) + 2 \frac{\delta t}{\delta x} (1 + \varepsilon^2)\right)(\phi_1 - \phi_2)^2 + (\varepsilon - i)(\phi_1 + \phi_2)^2\right]} \,.
\end{align}
It is readily checked that
\begin{equation}
\lim_{\varepsilon \to 0^+}\int d\phi_1 \, d\phi_2 \, \Psi^*(\phi_1, \phi_2) \Psi(\phi_1, \phi_2) = 1\,,
\end{equation}
which is in fact a consequence of $\mathcal{K}$ being an isometry.  Indeed, the order of limits is important since
\begin{equation}
\int d\phi_1 \, d\phi_2 \left(\lim_{\varepsilon \to 0^+}\Psi^*(\phi_1, \phi_2)\right) \left(\lim_{\varepsilon \to 0^+} \Psi(\phi_1, \phi_2) \right) = 0\,.
\end{equation}
In particular, $\lim_{\varepsilon \to 0^+} \Psi(\phi_1, \phi_2) = 0$ which is of course not a normalizable wavefunction.

Clearly this dependence on $\varepsilon$ augments what we mean by a quantum wavefunction in the present setting.  Let us give a proposal for how to treat this.  In section~\ref{Subsec:pathintegral1} we said that in the context of a free scalar on a triangle, $\mathcal{K}$ is a map $\mathcal{K} : \mathcal{H}_0 \to \mathcal{H}_1 \otimes \mathcal{H}_2$.  But what exactly are these Hilbert spaces $\mathcal{H}_i$?  Our initial inclination would be to say that they are all $L^2(\mathbb{R})$, but it will be convenient to refine this slightly.

In single particle quantum mechanics, we usually think of the Hilbert space as being $L^2(\mathbb{R})$, but a more refined approach is to consider a rigged Hilbert space, also called a Gelfand triple; see~\cite{de2005role} for a nice discussion.  We briefly recall the main ingredients.

The problem with the single-particle quantum mechanics Hilbert space being $L^2(\mathbb{R})$ is that the $\hat{x}$ and $\hat{p}$ operators and polynomials thereof do not map $L^2(\mathbb{R})$ to itself.  For instance, $\psi(x) = \frac{1}{\sqrt{\pi}} \frac{1}{\sqrt{1 + x^2}}$ is in $L^2(\mathbb{R})$ but acting $\hat{x}$ on the state gives us $x\,\psi(x)$ which is not in $L^2(\mathbb{R})$.  To ameliorate this issue, we consider Schwartz class functions $\mathcal{S}(\mathbb{R})$ which are a subset of $L^2(\mathbb{R})$ with the property that any operator which is a finite order polynomial in $\hat{x}$'s and $\hat{p}$'s maps $\mathcal{S}(\mathbb{R})$ to itself.  We can then say that physical wavefunctions live in $\mathcal{S}(\mathbb{R})$.  However, eigenfunctions of $\hat{x}$ and $\hat{p}$ (and also eigenfunctions of polynomial combination thereof) need not be in either $\mathcal{S}(\mathbb{R})$ or the larger space $L^2(\mathbb{R})$.  For example we can think of $\delta(x)$ or $e^{i p x}$.  To accommodate these wavefunctions, we consider $\mathcal{S}^\times(\mathbb{R})$ which is the space of antilinear functionals over $\mathcal{S}(\mathbb{R})$.  (If we considered linear functionals over $\mathcal{S}(\mathbb{R})$ then we would have a space of bras; we want a space a kets and so consider antilinear functionals.)  Then our rigged Hilbert space is
\begin{equation}
\mathcal{S}(\mathbb{R}) \subset L^2(\mathbb{R}) \subset \mathcal{S}^\times(\mathbb{R})\,,
\end{equation} 
where as we said before the physical states live in $\mathcal{S}(\mathbb{R})$, and eigenfunctions of operators can live in $\mathcal{S}^\times(\mathbb{R})$.

Returning to our problem of interest, we might guess that at the very least $\mathcal{K}$ maps $\mathcal{S}(\mathbb{R}) \to \mathcal{S}(\mathbb{R}^2)$, but this is not quite right.  In fact, $\mathcal{K}$ maps $\mathcal{S}(\mathbb{R})$ to Schwartz class functions from $\mathbb{R}^2 \to \mathbb{C}$ which also have a dependence on $\varepsilon$.  To notate this, it is convenient to expand our function space notation so that for instance $\mathcal{S}(\mathbb{R}^2, \mathbb{C})$ denotes Schwartz class functions with domain $\mathbb{R}^2$ and codomain $\mathbb{C}$.  We want to expand our codomain space $\mathbb{C}$ to accommodate $\varepsilon$-dependence, and we call this augmented space ${}^*\mathbb{C}$.  So then we can write $\mathcal{K} : \mathcal{S}(\mathbb{R}, \mathbb{C}) \to \mathcal{S}(\mathbb{R}^2, {}^*\mathbb{C})$, or if we also allow the initial wavefunctions to have $\varepsilon$ dependence then we have $\mathcal{K} : \mathcal{S}(\mathbb{R}, {}^*\mathbb{C}) \to \mathcal{S}^{\times}(\mathbb{R}^2, {}^*\mathbb{C})$.

The notation ${}^*\mathbb{R}$ in fact denotes the hyperreals and likewise ${}^*\mathbb{C}$ denotes the hypercomplex numbers, both central objects in the theory of non-standard analysis~\cite{robinson1974non, loeb2000nonstandard}.  Non-standard analysis is an alternative formulation of real analysis which extends the real numbers (and complex numbers) to furnish infinitesimals and infinite numbers, providing a rigorous foundation for the original conception of calculus.  While this framework is somewhat high-powered, it appears appropriate to use it in our context as a bookkeeping device.  The only notation we will need, besides the ${}^*\mathbb{C}$ symbol, is `$\text{st}$'.  This denotes the `standard' part of a real number; for instance $\text{st}(x + \varepsilon \, y) = x$, and for a smooth function $f(x,y)$ we have $\text{st}(f(x, \varepsilon)) = \lim_{\varepsilon \to 0^+} f(x,\varepsilon)$.  See~\cite{robinson1974non, loeb2000nonstandard} for a complete discussion.

With the above in mind, what inner product should we choose for $L^2(\mathbb{R}, {}^*\mathbb{C})$ or $L^2(\mathbb{R}^2, {}^*\mathbb{C})$ (which induces an inner product on $\mathcal{S}(\mathbb{R}, {}^*\mathbb{C})$ or $\mathcal{S}(\mathbb{R}^2, {}^*\mathbb{C})$\,)?  For two functions $\psi(x,\varepsilon)$ and $\psi'(y,\varepsilon)$ on $L^2(\mathbb{R}, {}^*\mathbb{C})$, we choose the semi-inner product
\begin{equation}
\label{E:newL2}
\langle \psi, \psi' \rangle := \text{st}\,\int dx \, \psi^*(x,\varepsilon) \psi'(x,\varepsilon)\,,
\end{equation}
and we have analogous inner products on the other function spaces.  Note that the we are essentially taking the $\varepsilon \to 0^+$ limit on the outside, which comports with our discussion above about orders of limits.  The above induces a semi-norm in the usual way.  We use the prefix `semi' since the $\varepsilon$ part of a wavefunction will essentially be taken to zero when we compute inner products and norms, e.g.~$\varepsilon \, \psi(x)$ has zero norm.  When we speak of the nullspace of $\mathcal{K}^\dagger$ or $\mathcal{K} \mathcal{K}^\dagger$, what we operationally mean is the space of vectors which are mapped to zero norm states in the sense of~\eqref{E:newL2}

The above considerations suggest we should work with the augmented rigged Hilbert spaces
\begin{equation}
\mathcal{S}(\mathbb{R}, {}^*\mathbb{C}) \subset L^2(\mathbb{R}, {}^*\mathbb{C}) \subset \mathcal{S}^\times(\mathbb{R}, {}^*\mathbb{C})\,, \qquad \mathcal{S}(\mathbb{R}^2, {}^*\mathbb{C}) \subset L^2(\mathbb{R}^2, {}^*\mathbb{C}) \subset \mathcal{S}^\times(\mathbb{R}^2, {}^*\mathbb{C})\,,
\end{equation}
and our kernel may be viewed as implementing the maps
\begin{equation}
\label{E:mappingstoexplain1}
\mathcal{K} : \mathcal{S}(\mathbb{R}, {}^*\mathbb{C}) \to \mathcal{S}^{\times}(\mathbb{R}^2, {}^*\mathbb{C})\,, \qquad \mathcal{K}^\dagger : \mathcal{S}(\mathbb{R}^2, {}^*\mathbb{C}) \to \mathcal{S}^\times(\mathbb{R}^2, {}^*\mathbb{C})\,.
\end{equation}
Relatedly, we have
$\mathcal{K} \mathcal{K}^\dagger : \mathcal{S}(\mathbb{R}^2, \mathbb{C}) \to \mathcal{S}^\times(\mathbb{R}^2, {}^*\mathbb{C})$ or more generally
\begin{equation}
\mathcal{K} \mathcal{K}^\dagger : \mathcal{S}(\mathbb{R}^2, {}^*\mathbb{C}) \longrightarrow \mathcal{S}^\times(\mathbb{R}^2, {}^*\mathbb{C})\,.
\end{equation}
This is compatible with our finding in Appendix~\ref{App:isom1} where we found that normalizable states on sites $1$ and $2$ that are mapped to non-normalizable states via $\mathcal{K} \mathcal{K}^\dagger$.

% In summary, perhaps the most interesting output of this discussion is the proposal that we should allow physical states to have $\varepsilon$ dependence (i.e.~a `non-standard' part), and in doing so we should leverage the inner product in~\eqref{E:newL2}.

In summary, perhaps the most interesting output of this discussion is the proposal that we should allow some physical states to have $\varepsilon$ dependence (i.e.~a `non-standard' part), and in doing so we should leverage the inner product in~\eqref{E:newL2}.

\section{Isometries in interacting quantum field theories}
\label{App:isom2}

Our arguments about isometries in section~\ref{Subsec:pathintegral1} pertained to free fields, or more generally quadratic lattice actions.  We argued that
\begin{equation}
\label{E:Kprop2}
K(\vec{\phi}_F\,;\,\vec{\phi}_P) = C\,e^{i \left(\vec{\phi}_P \cdot Q \cdot \vec{\phi}_P + \vec{\phi}_F \cdot R \cdot \vec{\phi}_P + \vec{\phi}_F \cdot S \cdot \vec{\phi}_F \right) - \varepsilon \, T(\vec{\phi}_P, \vec{\phi}_F)}\,,
\end{equation}
where $T(\vec{\phi}_P, \vec{\phi}_F)$ is quadratic in the fields, insantiates an isometry when $|F| > |P|$.  We presently explain how our argument generalizes in the presence of non-quadratic interaction terms.

First it is useful to take a step back and understand how interaction terms interface with the FEM at the classical level.  A useful example is again the harmonic oscillator, with action $\int dt \left(\frac{1}{2}\dot{\phi}(t)^2 - \frac{1}{2}\phi(t)^2\right)$.  The interaction term here is simply the quadratic mass term, and using the triangular basis functions from subsection~\ref{Subsec:FEM1} we find the lattice action
\begin{equation}
S_{\text{lattice}}[\{\phi_i\}] = \sum_i a \left(\frac{1}{2} \left(\frac{\phi_{i+1} - \phi_i}{a}\right)^2 - \frac{1}{2} \cdot \frac{1}{3}\left(\phi_i^2 + \left(\frac{\phi_i + \phi_{i+1}}{2}\right)^2 + \left(\frac{\phi_i + \phi_{i-1}}{2}\right)^2\right) \right)\,.
\end{equation}
The first term in fact agrees with the usual FDM discretization of the time derivative, but the second block of terms is less familiar.  Using the FDM, we typically discretize $\int dt \, \phi(t)^2 \to \sum_i a \, \phi_i^2$ whereas with our present FEM discretization we have
\begin{equation}
\label{E:massterm1}
\int dt \, \phi(t)^2 \longrightarrow \int dt \left(\sum_k \phi_k \, b_k(t)\right)^2 = \sum_i a \, \frac{1}{3}\left(\phi_i^2 + \left(\frac{\phi_i + \phi_{i+1}}{2}\right)^2 + \left(\frac{\phi_i + \phi_{i-1}}{2}\right)^2\right)\,.
\end{equation}
At the classical level, this is an acceptable alternative, even though it couples the field at different times.  In the corresponding setting of a free massive scalar field, this coupling at different times will not be problematic for our path integral analysis of isometries since the lattice action will still be quadratic.

However, when we consider higher-order interactions in field theory, for example a $\phi^4$ term (we will be mostly interested in polynomials of the field instead of higher derivative terms), the kind of FEM discretization analogous to~\eqref{E:massterm1} yielding couplings between different time slices will pose technical challenges to our isometry analysis.  Shortly we will see why this is the case.  As such, it is desirable to have interaction terms which do not couple between different times.

In the FEM, most of the subtleties of convergence in the continuum limit are tied up with the derivative terms in the PDE or corresponding action functional (if one exists).  The approximation of non-derivative interaction terms is more flexible.  A standard approximation is as follows: when we have integrals of the form $\int dt \left(\sum_k \phi_k \, b_k(t)\right)^k$ for integer $k$, we perform a trapezoidal Riemann sum approximation of the integral with lattice scale $a$.  Recall that this approximation is $\int_{\text{trap.}} dt \, f(t) = \sum_i a \, \frac{f(i a) + f((i+1)a)}{2}$.  It is readily checked that
\begin{equation}
\int_{\text{trap.}} dt \left(\sum_k \phi_k \, b_k(t)\right)^k = \sum_i a \, \phi_i^k\,,
\end{equation}
namely we now have the more standard interaction terms which do not couple across multiple times.  This same trapezoidal approximation can be made with minimal modifications when we have boundary conditions.  Moreover, an appropriately generalized version works in the higher dimensional setting (i.e.~field theory setting) with triangular lattices.  There are analogous, albeit more sophisticated versions of this approximation when we have suitable basis functions that are not piecewise linear (for instance, piecewise quadratic).

For our purposes, let us consider a $d+1$ field theory with continuum action
\begin{align}
\label{E:Sinteracting1}
S[\phi(t)] = - \int d^{d+1} x \, \left(\frac{1}{2}\,\eta^{\mu \nu} \partial_\mu \phi \, \partial_\nu \phi + V(\phi) \right) + i \varepsilon  \int d^{d+1} x \, \left(\frac{1}{2}\,\delta^{\mu \nu} \partial_\mu \phi \, \partial_\nu \phi + V(\phi) \right) 
\end{align}
where we have instantiated an appropriate $i \varepsilon$ prescription.  For simplicity, let the action have $\mathbb{Z}_2$ symmetry so that $V(\phi)$ is a sum of even powers.  Then using the notation of~\eqref{E:Kprop2}, we let the lattice action restricted to two adjacent Cauchy slices take the form
\begin{align}
\label{E:latticenew1}
S(\vec{\phi}_P, \vec{\phi}_F) := \left(\vec{\phi}_P \cdot Q \cdot \vec{\phi}_P + \vec{\phi}_F \cdot R \cdot \vec{\phi}_P + \vec{\phi}_F \cdot S \cdot \vec{\phi}_F -(1 - i \varepsilon) \big(V(\vec{\phi}_P) + V(\vec{\phi}_F)\big)\right) + i \varepsilon \, T(\vec{\phi}_P, \vec{\phi}_F)
\end{align}
where $T(\vec{\phi}_P, \vec{\phi}_F)$ is again quadratic in the fields.  Here the potential terms $V(\vec{\phi}_P)$ and $V(\vec{\phi}_F)$ do not couple; we can regard this as having performed an appropriate trapezoidal Riemann sum approximation on the potential term in the action as per our discussion above, or we can simply view this as an approximation by fiat on top of the FEM (which is often how this is treated).

This kind of approximation has been recently leveraged in the setting of Euclidean quantum fields coupled to curved manifolds (see~\cite{Brower:2016moq} for an overview).  These works used an augmentation of the FEM, and discovered that in their setup it was prudent to incorporate certain local counterterms to achieve the desired continuum limit of the quantum theory.  Presumably similar counterterms should be incorporated into our Lorentzian setting, but they will not change the schematic form of the action in~\eqref{E:latticenew1}.  Since the schematic form is the only essential ingredient to our arguments which follow, we will not pursue the issue of counterterms here.

Proceeding with our analysis,~\eqref{E:latticenew1} tells us that our propagator between the two adjacent Cauchy slices should be
\begin{equation}
\label{E:Kprop3}
K(\vec{\phi}_F\,;\,\vec{\phi}_P) = C\,e^{i \left(\vec{\phi}_P \cdot Q \cdot \vec{\phi}_P + \vec{\phi}_F \cdot R \cdot \vec{\phi}_P + \vec{\phi}_F \cdot S \cdot \vec{\phi}_F - V(\vec{\phi}_P) - V(\vec{\phi}_F)\right) -\varepsilon \left(V(\vec{\phi}_P) + V(\vec{\phi}_F)\right) - \varepsilon \, T(\vec{\phi}_P, \vec{\phi}_F)}\,.
\end{equation}
Computing the past to future to past propagator, we find
\begin{align}
&\int d\vec{\phi}_F \, K^*(\vec{\phi}_F\,;\,\vec{\phi}_P')  \,  K(\vec{\phi}_F\,;\,\vec{\phi}_P) \nonumber \\
& \qquad =  |C|^2\,e^{i \left(\vec{\phi}_P \cdot Q \cdot \vec{\phi}_P - \vec{\phi}_P' \cdot Q \cdot \vec{\phi}_P' + \vec{\phi}_F \cdot R \cdot (\vec{\phi}_P - \vec{\phi}_P') \right) - i \left(V(\vec{\phi}_P) - V(\vec{\phi}_P')\right)} \nonumber \\
& \qquad \qquad \qquad \qquad \qquad \qquad \times e^{- \varepsilon \left(V(\vec{\phi}_P) + V(\vec{\phi}_F)\right) - \varepsilon \left(V(\vec{\phi}_P') + V(\vec{\phi}_F)\right) - \varepsilon \, T(\vec{\phi}_P, \vec{\phi}_F) - \varepsilon \, T(\vec{\phi}_P', \vec{\phi}_F)}\,.
\end{align}
Then following the same approach as in section~\ref{Subsec:pathintegral1} for $|F| > |P|$, we are left with
\begin{equation}
\label{E:newdelta1}
e^{- i \left(V(\vec{\phi}_P) - V(\vec{\phi}_P')\right)} \delta^{|P|}(\vec{\phi}_P - \vec{\phi}_P') = \delta^{|P|}(\vec{\phi}_P - \vec{\phi}_P')\,.
\end{equation}
A difference with our previous analysis is that an additional factor $e^{- 2 \varepsilon \, V(\vec{\phi}_F)}$ participates in the $\vec{\phi}_F$ integral, but this only serves as an additional contribution to the $\varepsilon$-regularizer $e^{- \varepsilon \, T(\vec{\phi}_P, \vec{\phi}_F) - \varepsilon \, T(\vec{\phi}_P', \vec{\phi}_F)}$ of the $\delta$ distribution.  Here~\eqref{E:newdelta1} indeed shows us that $\mathcal{K}^\dagger \mathcal{K}$ is the identity.

The computation of the future to past to future propagator works in a similar fashion.  Then following the same argument as in section~\ref{Subsec:pathintegral1} establishes that $\mathcal{K} \mathcal{K}^\dagger$ is not the identity, and so $\mathcal{K}$ is an isometry that is non-unitary.

We end with several remarks.  In our above analysis it was convenient that the interaction terms did not couple between different times.  This enabled cancellations which allowed us to essentially recapitulate our same isometry analysis from section~\ref{Subsec:pathintegral1}.  If the interactions coupled between different times, it would not be clear how to proceed.  However, if such interactions between different times were due to artifacts of the lattice approximation of ordinary $V(\phi)$ potential terms (such as by using the standard FEM approximation and not doing the trapezoidal Riemann sum augmentation), then we are inclined to believe such interactions are ultimately benign in the quantum analysis.  One possibility is that the appropriate path integral measure in this case is not a flat one, which would make $C$ a function of fields at the level of our propagator analysis.  On the other hand, if interactions between different times were induced by higher derivative terms in the action then it is less clear how to proceed, even for establishing unitarity in more standard lattice field theory settings.  A further complicating issue is that higher derivative terms in Poincar\'{e}-invariant actions can lead to superluminal signaling even at the classical level; presumably we would want to avoid the combinations of higher-derivative couplings which do this before attempting to analyze unitary or isometry properties of a higher-derivative quantum field theory.

\bibliography{refs}
\bibliographystyle{JHEP}

\end{document}